\documentclass{article}

\usepackage{arxiv}

\usepackage[utf8]{inputenc} 
\usepackage[T1]{fontenc}    
\usepackage{hyperref}       
\usepackage{url}            
\usepackage{booktabs}       
\usepackage{amsfonts}       
\usepackage{nicefrac}       
\usepackage{microtype}      
\usepackage{graphicx}
\usepackage{natbib}
\usepackage{doi}
\usepackage{rotating}

\hypersetup{
    colorlinks=false,   
    linkcolor=black,   
    urlcolor=blue,     
    citecolor=black,   
    pdfborder={0 0 0}, 
}

\title{Automatic Sensor-free Affect Detection: \\A Systematic Literature Review}

\author{ \href{https://orcid.org/0000-0002-8510-4516}{\includegraphics[scale=0.06]{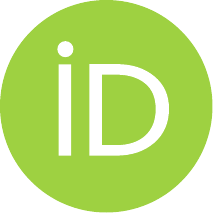}\hspace{1mm}Felipe~de~Morais} \\
	UNISINOS\\
	\texttt{felipedemoraisfdm@gmail.com} \\
	\And
	\href{https://orcid.org/0000-0001-8675-6287}{\includegraphics[scale=0.06]{images/orcid.pdf}\hspace{1mm}Diógines D'Ávila Goldoni} \\
	UNISINOS\\
	\texttt{stariate@ee.mount-sheikh.edu} \\
    \And
	\href{https://orcid.org/0000-0002-6017-8340}{\includegraphics[scale=0.06]{images/orcid.pdf}\hspace{1mm}Tiago Kautzmann} \\
	IFSul\\
	\texttt{tkautzmann@gmail.com} \\
 \And
	\hspace{1mm}Rodrigo Rodrigues da Silva \\
	UNISINOS\\
	\texttt{rodrigo\_lrsilva@hotmail.com} \\
 \And
	\href{https://orcid.org/0000-0002-2933-1052}{\includegraphics[scale=0.06]{images/orcid.pdf}\hspace{1mm}Patricia A. Jaques} \\
	PPGInf/UFPR \\
    PPGC/UFPEL \\
	\texttt{patricia@inf.ufpr.br} \\
}

\renewcommand{\shorttitle}{Automatic Sensor-free Affect Detection}

\hypersetup{
pdftitle={Automatic Sensor-free Affect Detection: \\A Systematic Literature Review},
pdfsubject={q-bio.NC, q-bio.QM},
pdfauthor={Felipe de Morais, Diógines Goldoni, Tiago Kautzmann, Rodrigo da Silva, Patricia A. Jaques},
pdfkeywords={sensor-free affect detection, systematic literature review, emotional learning environments, emotion detection, computer-based learning environments},
}

\begin{document}
\maketitle

\begin{abstract}
	Emotions and other affective states play a pivotal role in cognition and, consequently, the learning process. It is well-established that computer-based learning environments (CBLEs) that can detect and adapt to students' affective states can enhance learning outcomes. However, practical constraints often pose challenges to the deployment of sensor-based affect detection in CBLEs, particularly for large-scale or long-term applications. As a result, sensor-free affect detection, which exclusively relies on logs of students' interactions with CBLEs, emerges as a compelling alternative. This paper provides a comprehensive literature review on sensor-free affect detection. It delves into the most frequently identified affective states, the methodologies and techniques employed for sensor development, the defining attributes of CBLEs and data samples, as well as key research trends. Despite the field's evident maturity, demonstrated by the consistent performance of the models and the application of advanced machine learning techniques, there is ample scope for future research. Potential areas for further exploration include enhancing the performance of sensor-free detection models, amassing more samples of underrepresented emotions, and identifying additional emotions. There is also a need to refine model development practices and methods. This could involve comparing the accuracy of various data collection techniques, determining the optimal granularity of duration, establishing a shared database of action logs and emotion labels, and making the source code of these models publicly accessible. Future research should also prioritize the integration of models into CBLEs for real-time detection, the provision of meaningful interventions based on detected emotions, and a deeper understanding of the impact of emotions on learning.
\end{abstract}


\section{Introduction}

Various cognitive and affective factors play pivotal roles in facilitating or hindering learning processes in students. A confluence of elements such as affect, motivation, and meta-cognition influences cognition, thus impacting learning outcomes \citep{azevedo2013international}.

Of particular importance are the roles played by emotions and affective states in the learning process. A student's emotions can significantly influence their learning by governing attention, motivation, cognition, modeling strategies, and self-regulation \citep{pekrun2014emotions}. Emotions such as joy, engagement, and curiosity, for example, have been shown to enhance physical, social, intellectual, and creative aspects, thus contributing positively to skill development \citep{fredrickson1998good}. Conversely, frustration and boredom may instigate a cognitive imbalance that hampers the learning process \citep{graesser2011theoretical}. Interestingly, confusion presents a dual facet—when aptly regulated and resolved, it can improve learning, but if left unchecked, it can impair it \citep{dmello2014confusion}.

As our understanding of the affective dimensions of learning grows, the application of this knowledge in Computer-Based Learning Environments (CBLEs) becomes paramount. Research has shown that CBLEs, which can sense and adapt to the affective states of students, facilitate learning \citep{Arroyo2016,dmello2010, Litman2014}. However, for these systems to be 'affect-aware,' they must successfully detect and interpret emotions, which is an ongoing challenge in the field.

Many methods, including facial expressions, voice synthesis, and posture analysis, have been proposed to understand learners' affect \citep{calvo2010affect}. Nonetheless, these methods frequently rely on intrusive and expensive devices like cameras or sensors, limiting their practicality for large-scale or long-term use \citep{baker2012towards}.

Alternatively, monitoring students' interactions with the system through the Graphical User Interface (GUI) offers a more organic and less invasive means of affect detection \citep{salmeron2014evaluation,harley2016measuring}. These interactions produce logs that encapsulate valuable data about system usage, enabling researchers to detect students' affect in CBLEs by mining these logs \citep{paquette2016sensor}. Referred to as "sensor-free affect detection," this approach avoids the need for specific sensors \citep{baker2012towards}.

The implementation of sensor-free detection algorithms entails the use of machine learning techniques and statistical analysis. This process initiates with the collection of data from students' interactions within the learning environment, as well as the concurrent annotation of their emotions. Subsequently, the data undergoes cleaning and preprocessing to discard noise or irrelevant information and is synchronized with the corresponding emotional annotations for the training phase. This preprocessed data is then fed into a machine learning model that is trained to recognize patterns associated with specific emotions. The model is continuously tested and refined until it can accurately detect the desired emotion, thereby eliminating the need to provide emotion labels.

Sensor-free affect detection offers several advantages. Firstly, it provides a non-intrusive method to understand students' emotional states—an essential facet of personalized learning. By analyzing interaction patterns, these algorithms can infer emotional states, circumventing the need for intrusive sensors or cumbersome self-reports. Secondly, these systems can operate at scale, furnishing insights for large student populations simultaneously—a boon for online learning platforms. Thirdly, the passive collection of data from existing digital traces reduces the burden on students and educators. When deployed and integrated within a learning environment, these systems can provide real-time feedback, enabling timely intervention when students struggle and thus enhancing the learning experience.

Despite the field's considerable trajectory, with published works dating back to 2011, and growing interest among researchers, there appears to be a lack of comprehensive reviews on sensor-free affect detection. Therefore, this paper aims to synthesize the current knowledge on affect detection devoid of sensors.  The review employs a methodology based on the five-step process proposed by \cite{petersen2008systematic}: \textit{i}) defining the research questions, \textit{ii}) conducting a search for primary studies, \textit{iii}) screening the papers, \textit{iv}) assigning keywords to abstracts, and \textit{v}) extracting and mapping the data. 


\section{Affect in learning}
\label{sec:affect-learning}

    This article investigates affect detection within CBLEs. To provide a foundation, it is important to first define the constructs of `affect', `affective states', and `emotions' that form the basis of our exploration. We adopt Scherer's classification, as outlined in \cite{Scherer2000}, which considers emotions as one aspect of affective states, accompanied by moods and dispositions such as personality traits. Emotions, characterized by their relatively brief duration and high intensity, are typically elicited by specific incidents.

Scherer later introduced the term ``affective phenomena" instead of ``affective state," emphasizing the dynamic and process-oriented nature of these phenomena across multiple components \citep{Scherer2005}. Consequently, the term `affect" is used interchangeably with ``affective phenomena." This concept aligns with \cite{Russell2003}'s description of ``core affect" as a ``neurophysiological state consciously accessible as a straightforward, non-reflective feeling comprising an integral mix of hedonic (pleasure–displeasure) and arousal (sleepy–activated) values."

Importantly, not all affective phenomena associated with learning are classified as emotions. For instance, interest, while crucial to the learning experience, is often categorized as a motivational state rather than a conventional emotion \citep{hidi2006}. Given these nuances and following the approach of scholars in the field of sensor-free affect detection \citep{pardos2014affective,affectSequences2019Andres,baker2014extending,botelho2017improving}, we adopt the broader term ``affect," encompassing both emotions and other affective states.

Although researchers in sensor-free emotion detection typically aim to discern affect, emotions have received more extensive scrutiny in learning settings. Emotions commonly observed in learning environments are referred to as learning-centered emotions \citep{graesser2012emotions,graesser2014emotions} or academic emotions, as per the Control-Value Theory (CVT) \citep{pekrun2002academic,pekrun2006achievement}. Academic emotions can be categorized into groups such as achievement emotions, epistemic emotions, and others (e.g., social emotions like anger) \citep{pekrun2012academic}. Achievement emotions are related to learning activities (e.g., frustration, boredom) or their outcomes (e.g., pride, anxiety).

In conclusion, the CVT proposes that emotions are influenced by their appraisal antecedents. Specifically, two sets of appraisals are crucial in the arousal of achievement emotions: (1) the degree of control students have over the achievement activities and their outcomes, and (2) the subjective value of the activities and their outcomes (i.e., how students perceive the importance of these activities). Epistemic emotions may be triggered by the cognitive characteristics and processing of task-related information, including confusion, surprise, and interest. Emotions can also be classified based on their valence (positive or negative) and arousal (activating or deactivating), resulting in four emotional categories: positive activating (enjoyment, curiosity/interest), positive deactivating (relaxation), negative activating (anger/frustration, confusion, anxiety), and negative deactivating (boredom).

The theoretical model proposed by \cite{dmello2012dynamics}, a prominent model of emotions in computer-based learning environments, posits that cognitive disequilibrium, experienced by students as they grapple with complex learning tasks, leads to a state of confusion. As per the model, when managed effectively, this confusion can serve as a springboard to re-engagement and rich learning experiences. Conversely, unregulated confusion can spiral into frustration or boredom, both of which negatively correlate with learning outcomes. Evidence supporting this model came from a study involving undergraduate psychology students in the United States.

More recently, \cite{karumbaiah2018implications,karumbaiah2019case} expressed concerns about the misapplication of the L metric, a frequently used tool in the research on affect dynamics in CBLEs, potentially leading to faulty conclusions. As a corrective measure, \cite{karumbaiah2021re} reviewed those studies where the L metric had been misused by excluding self-transitions (transitions from an emotion to itself) from the data without proper statistical analysis. The authors then proceeded to re-analyze the data from these studies using a refined version of the L metric.

The data were separated based on the students' country of origin: either the United States or the Philippines. Separate models were developed for each country, along with a general model encompassing both. Interestingly, the general model indicated a single significant positive transition, from engagement to confusion. In the model specific to the Philippines, no significant positive transitions emerged, while the United States model revealed four: from engagement to both confusion and frustration, and from both confusion and boredom to engagement. Among these transitions, only two (engagement to confusion, and confusion to engagement) align with the ones suggested by \cite{dmello2012dynamics}.

Recently, \cite{morais2023} conducted a  study on affect dynamics, utilizing data from Brazilian elementary school students. The researchers employed  PAT2Math, an intelligent tutoring system designed to assist students in learning algebra. To observe and annotate the students' emotions, the EmAP-ML protocol was utilized. This  involved trained human coders analyzing recordings of the students' faces, along with audio and computer screen videos, to identify emotions such as confusion, frustration, boredom, and engagement \citep{morais2023}.

The findings of this study were consistent with those of \cite{karumbaiah2021re}, and no dynamic model for deep learning, as proposed by \cite{dmello2012dynamics}, could be identified. It is worth considering that the students in this study may have possessed a high level of background knowledge in the subject matter being taught by the intelligent tutoring system. Consequently, it is possible that these students did not find themselves in a situation where they were actively engaged in deep learning. 

\section{Sensor-free Affect Detection}
\label{sec:affect-detection}

This study aims to understand sensor-free detectors, which employ data derived solely from student interactions with the CBLE through input devices such as keyboards and mouse. These interactions may include a variety of actions such as keystroke count, typing speed, and mouse movements \citep{vea2016modeling}, and can extend to information extracted from the CBLE, like the number of errors a student commits on a task \citep{wang2015towards}, the frequency of help requests \citep{baker2012towards}, and achieved goals \citep{sabourin2011modeling}. These models, termed as ``sensor-free detectors," have exhibited promising results in the literature \citep{arroyo2009emotion,paquette2016sensor,baker2012towards,botelho2017improving}.

Primarily, sensor-free detectors are developed as classifier or regression models, utilizing a supervised learning approach. Here, student action log data is labeled with corresponding emotions and employed for model development (or training). During this phase, algorithms are designed to learn the action patterns associated with specific emotions from the logs, resulting in a model that can predict the emotion even when not explicitly provided.

Supervised learning falls under the broader umbrella of machine learning paradigms that are employed when available data consists of labeled examples - where each data point has a feature set and a corresponding label. Supervised learning algorithms strive to learn a function that maps these features (input) to their respective labels (output). Each instance for affect detectors consists of a student action (or series of actions) within the CBLE, forming the input, and the detected emotion, constituting the output.  

Predominantly, it is utilized for regression and classification problems. The goal of a regression task is to predict a real value, such as the intensity of an emotion, while a classification problem aims to categorize an input into one of multiple discrete classes, such as a specific detected emotion. The classification task may either be binary (determining whether or not the student is engaged, for example) or multi-class (determining if the student is engaged, confused, frustrated, or bored).

\begin{figure}[ht]%
    \centering
    \includegraphics[width=1\textwidth]{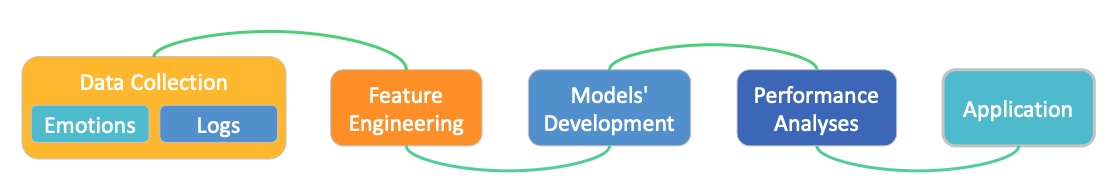}
    \caption{Phases of development of sensor-free affect detectors}
    \label{fig:method}
\end{figure}

Given that sensor-free affect detectors follow a supervised learning approach, their development process typically encompasses five main phases: data collection, feature engineering, model development, performance analysis, and application, as depicted in Figure \ref{fig:method}. Although this method is broadly applied across sensor-free affect detector development, the execution of each phase is contingent on the specific research objectives. 

\begin{itemize}
    \item \textbf{Data Collection:} This phase focuses on gathering the requisite information for the development of sensor-free detection models. It necessitates two types of data: action logs and emotions, which we categorize as sub-phases.
    \begin{itemize}
        \item \textbf{Logs:} Logs encapsulate student actions during CBLE interactions and are the models input. These actions can be varied, from clicks on the Graphical User Interface (GUI) to task completion, mistakes made, feedback received, and duration of each action, among others. The nature of the collected logs is determined by the specific CBLE in use.
        \item \textbf{Emotions:} Emotions serve as ground truth labels (output) for supervised machine learning training, crucial for the development of sensor-free affect detectors. These labels, synchronized with log data and calibrated to the observation window duration, can be collected using either a first-person or third-person approach or a mixed one. The first-person approach utilizes self-reporting, where students express their own emotions. This is typically achieved through periodic questionnaires, self-report scales \citep{wixon2014question}, or `emote-aloud' protocols \citep{porayska2013knowledge}, either administered at regular intervals or at the conclusion of a learning session. While providing direct access to a student's perceived emotions, periodic questionnaires can be interruptive and disruptive to the learning process. Additionally, due to social desirability bias or limited emotional self-awareness, students may not always accurately self-report their emotions. Contrarily, the third-person approach incorporates observer ratings, wherein trained observers, either co-located with students \citep{pedro2013predicting} or viewing recorded student videos \citep{morais2023}, label emotional states based on exhibited behavior. This method, potentially less intrusive, can offer a more objective perspective of a student's emotional state. Nevertheless, it may not always accurately capture the student's internal emotional state, particularly when certain emotions are not clearly manifested in observable behavior. Moreover, this method necessitates observer training.
    \end{itemize}
    These sub-phases may occur concurrently or sequentially. For methods that rely on crowdsourcing and log file annotation, log data collection is a prerequisite since it informs the analysis employed in emotion detection. Conversely, methods based on human observations or student self-reports can transpire alongside log collection. In essence, the synchronization between logs and emotions is a function of the approach adopted and refers to the accurate temporal and sequential identification of the emotion a student experiences when generating system interaction logs.

    \item \textbf{Feature Engineering:} After the data collection phase, an essential step in developing sensor-free affect detectors is feature engineering or data pre-processing. Feature engineering involves the transformation of raw data into features that better represent the underlying problem to the predictive models, resulting in improved model accuracy. In the context of sensor-free affect detectors, feature engineering might involve creating new variables such as the average time between actions, the number of errors per minute, or the frequency of specific action sequences. This process is crucial because the right features can simplify the learning process and improve the performance of the resulting models. It often involves domain expertise and a deep understanding of the data, as well as experimentation and iteration.
    
    \item \textbf{Model Development:} This phase comprises the selection of appropriate machine learning algorithms and the creation of emotion detection models through training. These models are trained with labeled data, i.e., data generated from logs and synchronized with emotion labels. Generally, in the realm of sensor-free affect detection, separate models are developed for each emotion to identify the best model for each emotion independently. The phase also includes data normalization, handling of missing data, class balancing, and feature selection for log representation.
    
    \item \textbf{Performance Analysis:} This phase aims to evaluate the efficacy of the developed models. Typically, model performance is gauged using various metrics that, in essence, indicate the likelihood of accurate student emotion classification based solely on log data. To assess model performance, unseen data is utilized for testing. Various validation methods are also applied to ensure the generalizability of model performance when deployed on different samples. 
    
    \item \textbf{Application:} This phase entails the utilization of developed models in experiments, analyses, or comparisons. While it is an optional phase—since some studies merely describe their approach to improving or developing different sensor-free detection models—many investigations employ these emotion detectors for various purposes. For instance, they may use recognized student emotions to predict learning session outcomes or to compare different emotion detection methods.

\end{itemize}

\section{Related Work}

This section presents an analysis of the relevant research surrounding Educational Data Mining (EDM), Learning Analytics (LA), and related subfields, with a specific focus on emotion detection methods in the educational context. The primary objective of this investigation was to identify any secondary or tertiary research related to sensor-free emotion detection in the field of education.

The initial strategy employed was a comprehensive search using Google Scholar with various combinations and variations of the terms \textit{sensor-free}, \textit{emotion}, and \textit{detection}. However, this initial search did not yield any similar research results. Most of the papers found were secondary and tertiary studies primarily focused on reviewing EDM or LA fields or exploring different types of emotion detection, such as sensor or text-based detection.

Subsequently, we initiated the search procedure for our Systematic Literature Review (SLR), which is detailed in Section \ref{sec:method}. During this stage, we established one specific exclusion criteria to classify papers as secondary and tertiary research. Upon completing the classification process, we compiled a list of all secondary and tertiary studies in the field, sourced from the most relevant digital libraries indexing papers in Computer Science, Affective Computing, Educational Data Mining, and Learning Analytics. These digital libraries include Web of Science, IEEE Xplore, Science Direct, ACM Digital Library, Engineering Village, ERIC, JEDM, Scopus, and Springer. It is noteworthy that this identical set of digital libraries was employed throughout the SLR process.

Next, we meticulously examined each paper identified as secondary and tertiary research from this comprehensive set. This analysis allowed us to select only those studies that were genuinely relevant to our work\footnote{The selection list of secondary and tertiary studies can be found here: \url{https://bit.ly/44R45eF}.}.

An overview of the EDM and LA fields can be found in \citep{tufekci2020educational}, which provides a comprehensive description of current research directions and remaining knowledge gaps. \cite{charitopoulos2020use} present a systematic review of literature on emotion analysis and sentiment analysis. They highlight that the field of emotion and affect detection is not yet mature enough to be fully integrated into automated procedures, suggesting that LA methods are currently more suitable. \cite{clow2013overview} offers an overview of LA with a focus on its educational role for teachers and related professionals, while \cite{romero2020educational} contextualize and define the field before discussing its state-of-the-art, supported by various examples.

Two systematic literature reviews focused on emotions in learning were also included. \cite{yadegaridehkordi2019affective} explore affect measurement in education using textual, visual, vocal, physiological, and multi-modal channels. \cite{mejbri2021trends} present a review of affective computing specifically in e-learning environments. The latter paper categorizes log-based emotion detection as a behavioral method but suggests the use of multi-modal emotion detection as a supplementary data source. Most papers reviewed employed unimodal data sources, with facial methods being the most common. Additionally, \cite{coto2021emotions} discuss the relationship between emotions and programming learning, presenting relevant tools, methods, and content, albeit not directly related to affect detection.

In the field of Affect Detection, \cite{garcia2017emotion} provide an overview of technologies and mention companies that offer emotion detection services, but no technologies specifically address sensor-free affect detection. \cite{d2017emotional} present an overview of the use of EDM and LA in affect research, with a particular focus on detection. Sensor-free methods are classified as affect detection methods based on interaction patterns. \cite{calvo2010affect} provide an overview of models, methods, and applications of affect detection without explicitly mentioning sensor-free affect detection. \cite{d2015review} and \cite{cernea2015survey} present surveys, with the former focusing on multi-modal affect detection and the latter discussing technologies and approaches used for emotion-enhanced interaction. \cite{d2017advanced} present 15 case studies in which researchers focused on detecting engagement, including both sensor-free and sensor-based methods. \cite{kappas2010smile} discuss the emergence of affective computing, its challenges, and the state-of-the-art in 2010 when the paper was written. The increasing popularity of affect detection is evident, with competitions dedicated to the field, such as the video-based detection competition highlighted by \cite{dhall2018emotiw}. \cite{bustos2021emotion} and \cite{acheampong2020text} provide an overview of advances, challenges, and opportunities related to text-based emotion detection, while \cite{kao2009towards} present a survey specifically focused on text-based emotion detection. 

The most comprehensive overview of sensor-free affect detection is provided by  \cite{baker2014interaction}. In their book chapter, the authors provide a comprehensive review of sensor-free affect detection, covering various aspects such as the learning environments targeted by these models, the methods used for collecting and annotating ground truth labels, the strategies employed for feature engineering, and the evaluation measures used by the studies. While this review offers a valuable insight into the field, it should be noted that it was published almost 10 years ago and do not address some of the specific research questions addressed in this paper.

Based on the research mentioned in this section, emotion detection emerges as a crucial aspect in the field of affective computing. The literature explores various strategies to deal with emotions. Recently, the EDM and LA fields have shown an increasing interest in utilizing CBLEs to gain insights into students' emotional states during the learning process. However, it was observed that only a limited number of papers discussed the state-of-the-art in the more specific field of sensor-free affect detection. Therefore, this study presents a systematic literature review to capture the recent developments in this field.

\section{Method} \label{sec:method}

This paper aims to give an overview of what scientists know about detecting emotions based only on students' actions in the CBLE. Therefore, we performed a systematic review method based on the five-step process proposed by \cite{petersen2008systematic}: \textit{i}) define the research questions, \textit{ii}) perform the search for primary studies, \textit{iii}) screen the papers, \textit{iv}) keyword of abstracts, and \textit{v}) extract and map the data.

\subsection{Definition of research questions}

According to our goal of presenting an overview of the sensor-free affect detection area, we have elaborated four Research Questions (RQ) to guide this work in collecting the most relevant research and summarizing the results obtained by the scientific community. For each RQ, we have also made sub-questions to look into each part of the area in more depth. 

\begin{itemize}
    \item \textbf{RQ1:} How are the emotions being considered by this type of detector?
\end{itemize}

RQ1 aims to identify the emotions detected and the theories on which authors have based their recognition and definition. It provides a summary of the types of emotions detected, how they are collected, at what granularity, based on which theories, and whether they have been taken into account with supplementary information.

\begin{itemize}
    \item \textbf{RQ2:} How are sensor-free affect detectors being developed?
\end{itemize}

The technical aspects of the machine learning methods used to create the sensor-free detectors are summarized in RQ2. It identifies the key algorithms and tools used to develop these models, the number of emotional labels, action protocols, and functions that are taken into consideration, the best results for detecting each emotion, and which measurements are used to identify them. It also identifies the most frequently applied features to detect each emotion and how they are collected and selected.

\begin{itemize}
    \item \textbf{RQ3:} For which scenarios are the sensor-free affect detectors being used?
\end{itemize}

RQ3 looks into the learning environment in which the emotion detectors were created. It covers the sample characteristics, the types of educational learning environments utilized, the learning domains and content used, and the works' objectives and purposes.

\begin{itemize}
    \item \textbf{RQ4:} What are the trends and main ideas for future research?
\end{itemize}

Finally, RQ4 examines the generalization performance of the produced detectors and how they may be utilized in production. It also investigates the trends and future work ideas in sensor-free affect detection. An overview of the present state of the field's research, including publication sources, authors, and years, is also provided.

\subsection{Search procedures}
We started by examining a number of initial papers provided by experts in the field. Then, we defined our research questions, and based on the keywords from the RQs, we created the search string. The search was conducted in the most relevant digital libraries that index papers on Computer Science, Affective Computing, Educational Data Mining, and Learning Analytics: Web of Science, IEEE Xplore, Science Direct, ACM Digital Library, Engineering Village, ERIC, JEDM, Scopus, and Springer. Different search strings were used according to the rules and syntax of each digital library. However, the core was based on \textit{``(sensor-free OR interaction-based OR log-based) AND (emotion OR affect) AND (detect OR predict OR infer)''}. The primary works were searched between November and December of 2021. All the metadata from the papers was exported from the digital libraries and imported into a shared spreadsheet. As the next step, the authors registered all the actions from the selection phases.

\subsection{Papers screening for inclusion and exclusion}
After importing all the data into a structured spreadsheet, we applied inclusion and exclusion criteria to filter the works. The Exclusion Criteria (EC) were:

\begin{itemize}
    \item (EC1) Duplicated papers;
    \item (EC2) Papers not written in English;
    \item (EC3) Papers published as secondary or tertiary studies; 
    \item (EC4) Non-peer-reviewed papers, books, proceedings, dissertations, thesis, summary, or short papers;
    \item (EC5) Papers focusing on affect detection based on text without comparing with log-based models (for a recent review on text-based emotion detection, see \citep{bustos2021emotion});
    \item (EC6) Papers focusing on sensors for detecting affect \footnote{This criterion includes sensors in mobile phones, for example, touchscreen pressure data, camera, etc.};
    \item (EC7) Papers focusing on affect detection in areas other than education;
    \item (EC8) Papers focusing on different areas than automatic affect detection;
    
\end{itemize}

The Inclusion Criteria (IC) were:

\begin{itemize}
    \item (IC1) Studies that detect emotions through logs generated by the student's interaction with the learning environment;
    \item (IC2) Studies comparing data-driven and sensor-driven approaches, that also provide details about the development of the models;
    \item (IC3) If several papers reported the same study, only the most recent was included;
\end{itemize}

\begin{figure}[ht]%
    \centering
    \includegraphics[width=1\textwidth]{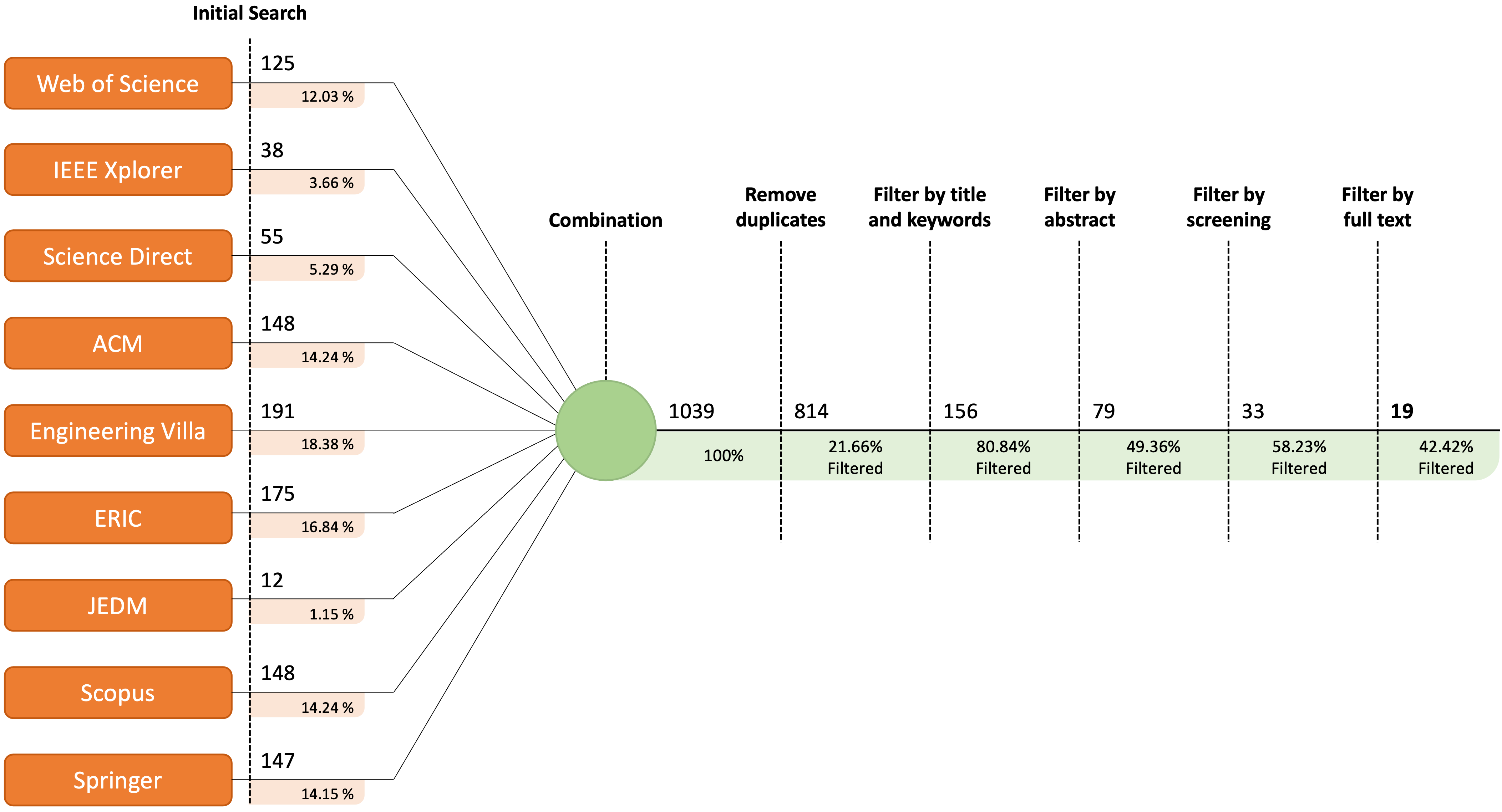}
    \caption{Papers selection process according to each digital library and the phases.}
    \label{fig:selection_process}
\end{figure}

The filtering process, guided by the application of the exclusion and inclusion criteria, is depicted in Figure \ref{fig:selection_process}. Our initial search across all digital libraries yielded 1,039 papers. Of these, 225 were removed due to duplication. The remaining 814 papers were randomly assigned to three reviewers (the authors) for initial screening based on the title and keywords, resulting in the rejection of 658 papers. The reviewers then assessed the abstracts of the remaining 156 papers, eliminating another 77. 
The full texts of the remaining 79 articles were screened, with 46 more papers being excluded due to insufficient information about the development of the emotion detection models. The reviewers conducted a thorough review of the final 33 articles, resulting in the elimination of 14 more. 
Specifically, 3 papers were excluded as they were non-peer-reviewed papers, books, proceedings, dissertations, thesis, summaries, or short papers (EC4), 3 papers were excluded for focusing on sensor-based affect detection (EC6), 1 paper was excluded for focusing on affect detection outside the realm of education (EC7), and 7 papers were excluded for focusing on areas other than automatic affect detection (EC8). Ultimately, 19 papers were included in the final selection, as listed in Table \ref{tab:selected_works}. 
The complete list of selected and eliminated papers at each phase, along with the corresponding criteria, can be found in the external worksheet at \url{https://bit.ly/44nJGhC}.

\begin{sidewaystable}
\begin{center}
\begin{minipage}{600pt}
\caption{Final set of selected papers from this Systematic Literature Review.}\label{tab:selected_works}%
\begin{tabular}{@{}lll@{}}
\toprule
\textbf{ID} & \textbf{Paper Title} & \textbf{Reference} \\

\midrule
P1 & \begin{tabular}[c]{@{}l@{}}Accuracy vs. Availability Heuristic in Multimodal Affect Detection \\ in the Wild\end{tabular} & \citep{bosch2015accuracy} \\
P2 & \begin{tabular}[c]{@{}l@{}}Sensor-Free or Sensor-Full: A Comparison of Data Modalities in \\ Multi-Channel Affect Detection\end{tabular} & \citep{paquette2016sensor} \\
P3 & \begin{tabular}[c]{@{}l@{}}Generalization performance of Sensor-Free Affect Detection Models in a \\ Longitudinal Dataset of Tens of Thousands of Students\end{tabular} & \citep{jensen2019generalizability} \\
P4 & \begin{tabular}[c]{@{}l@{}}A Comparison of Video-Based and Interaction-Based Affect \\ Detectors in Physics Playground\end{tabular} & \citep{kai2015comparison} \\
P5 & \begin{tabular}[c]{@{}l@{}}Affect Detection in Home-Based Educational Software for \\ Young Children\end{tabular} & \citep{smeets2019affect} \\
P6 & \begin{tabular}[c]{@{}l@{}}Affective States and State Tests: Investigating How Affect and \\ Engagement during the School Year Predict End-of-Year \\ Learning Outcomes\end{tabular} & \citep{pardos2014affective} \\
P7 & \begin{tabular}[c]{@{}l@{}}Population Validity for Educational Data Mining Models: A Case \\ Study in Affect Detection\end{tabular} & \citep{ocumpaugh2014population} \\
P8 & \begin{tabular}[c]{@{}l@{}}Disengagement Detection in Online Learning: Validation Studies \\ and Perspectives\end{tabular} & \citep{cocea2010disengagement} \\
P9 & \begin{tabular}[c]{@{}l@{}}Exploring the Effect of Student Confusion in Massive Open \\ Online Courses\end{tabular} & \citep{yang2016exploring} \\
P10 & \begin{tabular}[c]{@{}l@{}}Expert Feature-Engineering vs. Deep Neural Networks: Which \\ Is Better for Sensor-Free Affect Detection?\end{tabular} & \citep{jiang2018expert} \\
P11 & \begin{tabular}[c]{@{}l@{}}Detecting and Addressing Frustration in a Serious Game for \\ Military Training\end{tabular} & \citep{defalco2018detecting} \\
P12 & \begin{tabular}[c]{@{}l@{}}When the Question is Part of the Answer: Examining the Impact \\ of Emotion Self-reports on Student Emotion\end{tabular} & \citep{wixon2014question} \\
P13 & \begin{tabular}[c]{@{}l@{}}Improving Affect Detection in Game-Based Learning with \\ Multimodal Data Fusion\end{tabular} & \citep{henderson2020improving} \\
P14 & \begin{tabular}[c]{@{}l@{}}Sensor-Free Affect Detection for a Simulation-Based Science \\ Inquiry Learning Environment\end{tabular} & \citep{paquette2014sensor} \\
P15 & \begin{tabular}[c]{@{}l@{}}Time to Scale: Generalizable Affect Detection for Tens of \\ Thousands of Students across an Entire School Year\end{tabular} & \citep{hutt2019time} \\
P16 & \begin{tabular}[c]{@{}l@{}}Analysis and Prediction of Student Emotions While Doing \\ Programming Exercises\end{tabular} & \citep{tiam2019analysis} \\
P17 & Towards sensor-free affect detection in cognitive tutor algebra & \citep{baker2012towards} \\
P18 & \begin{tabular}[c]{@{}l@{}}Extending log-based affect detection to a multi-user virtual \\ environment for science\end{tabular} & \citep{baker2014extending} \\
P19 & Improving sensor-free affect detection using deep learning & \citep{botelho2017improving}  \\
\bottomrule
\end{tabular}
\end{minipage}
\end{center}
\end{sidewaystable}



\section{Results and Discussions}

This section presents the results and discussions according to each RQ.

\subsection{RQ1 - How are the emotions being considered by this type of detector?}

This RQ aims to uncover how emotions are being considered in the emotion detectors described in the selected works. More precisely, we divided this RQ to discover which emotions are being considered (RQ1.1), how they are obtained (RQ1.2), at what granularity they are obtained (RQ1.3), based on which theories (RQ1.4), and whether additional information is being considered (RQ1.5). These sub-questions are addressed in the following sections.

\subsubsection{RQ1.1 - Which emotions are being considered?}
\label{sec:rq1.1}

\begin{figure}[ht]%
    \centering
    \includegraphics[width=1\textwidth]{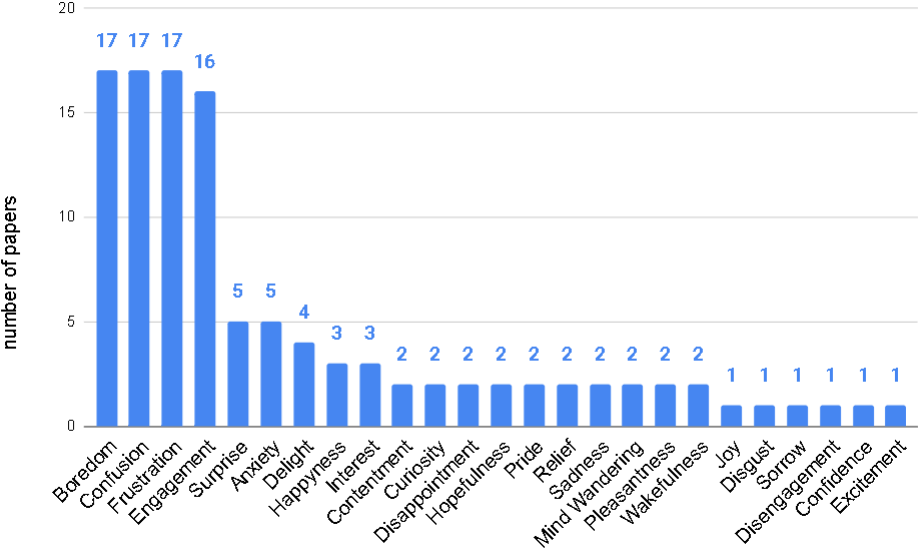}
    \caption{Emotions considered by the emotion detectors}
    \label{fig:which-emotions}
\end{figure}

To answer this question, we made a list of all the emotions considered by emotion detectors in the chosen works. So, we added how many times each emotion was considered. Figure \ref{fig:which-emotions} shows that boredom, confusion, frustration, and engagement are the most detected emotions in more than 84\% of the works. A possible justification for almost all studies considering these emotions in detection models is that these emotions are found to be more frequent in complex learning activities \citep{dmello2012dynamics} and during learning with technologies \citep{d2013selective,bosch2017affective,bosch2013emotions}.

\subsubsection{RQ1.2 - How are they obtained?}
\label{sec:rq1.2}

The selected works employed various methods to capture students' emotions using a CBLE. These methods can be categorized into third-person and first-person approaches (see Section \ref{sec:affect-detection}). The third-person approach includes annotations from human observers, crowdsourcing, and log-file annotation, while the first-person approach involves student self-reports.

Annotations from human observers, which is the most commonly used method, were utilized by several works \citep{bosch2015accuracy,paquette2016sensor,kai2015comparison,pardos2014affective,ocumpaugh2014population,jiang2018expert,defalco2018detecting,henderson2020improving,paquette2014sensor,baker2012towards,baker2014extending,botelho2017improving}. In this type of data collection, experts known as ``coders" observe students and annotate their emotions by analyzing their facial and physical expressions during the learning activity. This observation can be conducted through videos (offline) or directly in the classroom (online). Although other human-observer-based methods exist, such as the EmAP-ML protocol \citep{DeMorais2019}, the most widely referenced method in this category is the Baker-Rodrigo-Ocumpaugh Method and Protocol (BROMP) \citep{ocumpaugh2015baker}.

BROMP is an online protocol used for annotating students' emotions in real-time during their interactions with a CBLE. Trained human coders observe students using the CBLE in the classroom and record their unique emotions at 20-second intervals, following a round-robin fashion where each student is observed in sequence. If a student exhibits multiple emotions, the coder annotates the first one manifested and then proceeds to the next student. The protocol provides guidelines and predefined categories for labeling emotions and learning behaviors. According to the 2015 coding manual by \cite{ocumpaugh2015baker}, the most prevalent coding scheme, known as the ``PSLC" scheme, includes specific orientations for observers to collect five classes of learning behaviors (On-task, On-task Conversation, Off-task, Gaming the system, and Other for any other learning behavior) and five learning-centered emotions (Boredom, Confusion, Frustration, Engaged Concentration, and Other for any other emotion). These emotions and behaviors are identified by observing the students' facial expressions, physical gestures, and other behavioral cues exhibited during the learning activities. The coders undergo specific training to ensure their ability to accurately recognize and interpret students' emotional states based on the provided guidelines. The training typically involves familiarizing the coders with the CBLE, practicing emotion recognition techniques, and establishing inter-rater reliability to enhance consistency in the annotation process.

The second most commonly used method, student self-reports, was employed by \cite{jensen2019generalizability}, \cite{wixon2014question}, \cite{hutt2019time}, \cite{smeets2019affect}, and \cite{tiam2019analysis}. This method requires students to report their emotions during learning activities. For instance, the emote-aloud method \citep{porayska2013knowledge} involves students verbally expressing their emotions while interacting with the CBLE. Another approach is the free-response method \citep{d2006predicting}, in which students  use specific affect terms (e.g., happy), valence terms (e.g., slightly confused), or arousal terms (e.g., very active) to express their emotions in their own words.

The other two categories, crowdsourcing and log-file annotation, were each utilized by only one study. \cite{yang2016exploring} employed the crowdsourcing method, which involved a large group of paid participants annotating the students' confusion levels while reading their posts from the course forum. In this method, five coders assigned a confusion level to each sample, and the outcomes were determined based on the average or number of votes. On the other hand, log-file annotation, as applied by \citep{cocea2010disengagement}, involved retrospective analysis of student compilation logs in the system. Coders annotated students' emotions based on their perceptions of the data.

\subsubsection{RQ1.3 - At which grain level are the emotions obtained?}
\label{sec:req1.3}

The aim of this RQ is to determine the duration of the observation window for annotating and capturing students' actions within the system. In the literature, the term ``grain level" has been used to define this time period for obtaining emotions. Based on the examined papers, this time period can vary from seconds to weeks, depending on the approach used to collect emotions.

In the study by \cite{tiam2019analysis}, students' self-reports were annotated with variable time intervals, with an average of 17 seconds. Additionally, the collection process allowed the same student to make multiple annotations consecutively. For works that followed the BROMP protocol, each emotion was associated with a 20-second window \citep{bosch2015accuracy,paquette2016sensor,kai2015comparison,pardos2014affective,ocumpaugh2014population,jiang2018expert,defalco2018detecting,henderson2020improving,paquette2014sensor,baker2012towards,baker2014extending,botelho2017improving}. However, BROMP coders analyzed students in a round-robin fashion, meaning that the time between consecutive emotion labels could be minutes apart.

In \cite{wixon2014question}'s study, the system prompted students to identify their current emotions, with the pop-up appearing every 5 to 7 minutes. Similarly, \cite{smeets2019affect} requested students to self-report their emotions based on the same strategy, but with a maximum of three pop-ups per week. \cite{jensen2019generalizability} and \cite{hutt2019time} also employed a pseudo-randomly triggered pop-up strategy based on students' platform activities on their studies which aim was to investigate the generazibility of detection models across different student populations using data from 69,174
students over an entire
school year. These studies allowed for one pop-up every two weeks. In all of these methods, the recorded emotion was considered as a single instance, and neither the duration of the emotion nor the order in which students experienced it were recorded.

In \cite{cocea2010disengagement}'s study, the collection session was divided into 10-minute intervals, with each interval receiving a single label regarding students' engagement. The sessions were viewed sequentially, enabling the capture of a sequence of students' emotions. \cite{yang2016exploring} collected emotions from the class forum, with each post being labeled with an indication of the confusion level.

\subsubsection{RQ1.4 - In which works or theories are they based?}

This RQ aims to identify the psychological theories or works on emotions that the selected articles relied on to determine which emotions to detect. Our analysis revealed that the selection of emotions for detection was typically validated based on either the emotions addressed in the protocol used to collect ground truth labels or earlier research in emotion detection. A summary of this information is presented in Table \ref{tab:tabela_theories}. An overview of the main theories utilized is provided in Section \ref{sec:affect-learning}.

Most selected works \citep{paquette2016sensor,jiang2018expert,defalco2018detecting,henderson2020improving,botelho2017improving,baker2014extending} relied on the coding scheme from the BROMP protocol. 
As described in Section \ref{sec:rq1.2}, the BROMP protocol is a method designed to collect labels in the classroom by trained human coders who record students' emotions and behaviors based on a predetermined coding schema.  Hence, the previously cited efforts based on BROMP capture the five emotions mentioned in this protocol or a subset thereof.  In addition, the papers \citep{smeets2019affect,paquette2014sensor,baker2012towards,cocea2010disengagement} were based on prior research conducted by Baker and colleagues \citep{baker2010better,baker2012towards,botelho2017improving} and consequently on the BROMP protocol.

Several works selected their emotions based on educational frameworks or theories of emotions. The emotion selection in the works \citep{jensen2019generalizability,hutt2019time} is based on the Control-Value theory of emotions in education by Pekrun
\citep{pekrun2002positive,pekrun2007emotions,pekrun2012academic}.
\cite{wixon2014question} is based on the affective model of Kort and Picard \citep{kort2001affective}. Others  have cited  \cite{graesser2011theoretical} model of emotions in deep learning in computer-based learning settings.  
For instance, the work of \cite{yang2016exploring} is based on prior studies conducted by Baker and the Graesser and D'Mello model \citep{lehman2008you,lehman2012interventions,dmello2014confusion,dmello2012dynamics,pardos2014affective,lee2011exploring}. The study of \cite{pardos2014affective} is based on several previous works, including those of Baker and Graesser and D'Mello \citep{aleven2004toward,baker2007modeling,baker2010better,baker2012towards,cocea2009impact,craig2004affect,lee2011exploring,lehman2012confusion,rodrigo2009affective}.

Some efforts, however, relied their choice of emotions on prior research in emotion detection.
 \cite{bosch2015accuracy} based their emotion selection on D'Mello's meta-analysis on emotions in learning technologies \citep{d2013selective} as well as on their personal observations of students during the first day of data collection. 
The study of \cite{kai2015comparison} is based on D'Mello's research on the bodily expression of affect \citep{dmello2011dynamical}.
 The work of \cite{ocumpaugh2014population} based their  choice of emotions on different works for each emotion: boredom \citep{csikszentmihalyi1990flow,miserandino1996children}, confusion \citep{craig2004affect,kort2001affective}, engagement \citep{csikszentmihalyi1990flow}, and frustration \citep{kort2001affective,patrick1993motivates}.
Lastly, the work of \cite{tiam2019analysis} made no mention of their decision about the emotions considered.

\begin{table}[htbp]
\centering
\resizebox{\textwidth}{!}{%
\begin{tabular}{|l|l|}
\hline
\textbf{Original Paper} & \textbf{Theories or works considered as foundations} \\ \hline
\begin{tabular}[c]{@{}l@{}}
\citep{paquette2016sensor} \\ 
\citep{jiang2018expert} \\ 
\citep{defalco2018detecting} \\
\citep{henderson2020improving} \\
\citep{botelho2017improving} \\
\citep{baker2014extending} \\
\citep{smeets2019affect} \\
\citep{paquette2014sensor} \\
\citep{baker2012towards} \\
\citep{cocea2010disengagement} \\
\citep{pardos2014affective}
\end{tabular} & 
\begin{tabular}[c]{@{}l@{}}
BROMP protocol \citep{ocumpaugh2015baker}
\end{tabular} \\ \hline
\begin{tabular}[c]{@{}l@{}}
\citep{jensen2019generalizability} \\
\citep{hutt2019time}
\end{tabular} & 
\begin{tabular}[c]{@{}l@{}}
Control-Value theory \citep{pekrun2012academic}
\end{tabular} \\ \hline
\citep{wixon2014question} & 
Affective model of Kort and Picard \citep{kort2001affective} \\ \hline
\begin{tabular}[c]{@{}l@{}}
\citep{yang2016exploring} \\
\citep{pardos2014affective}
\end{tabular} & 
Graesser and D'Mello's model \citep{graesser2011theoretical} \\ \hline
\citep{bosch2015accuracy} & 
D'Mello's meta-analysis \citep{d2013selective} \\ \hline
\citep{kai2015comparison} & 
D'Mello's research \citep{dmello2011dynamical} \\ \hline
\citep{ocumpaugh2014population} & 
Different works for each emotion: \\ & - boredom \citep{csikszentmihalyi1990flow,miserandino1996children}, \\ & - confusion \citep{craig2004affect,kort2001affective}, \\ & - engagement \citep{csikszentmihalyi1990flow}, \\ & - frustration \citep{kort2001affective,patrick1993motivates} \\ \hline
\citep{tiam2019analysis} & 
Not mentioned \\ \hline
\end{tabular}%
}
\caption{Theories or works on emotions used as foundations for the analyzed papers}
\label{tab:tabela_theories}
\end{table}

\subsubsection{RQ1.5 - There is additional information considered with the emotions?}
\label{sec:rq1.5}

Research in the field of learning has demonstrated that emotions play a significant role in influencing subsequent emotional experiences \citep{dmello2012dynamics,dmello2014confusion,bosch2017affective,sinclair2018changes}. However, it is important to acknowledge that various factors can also impact the transition of students' emotions during the learning process. These factors include gender, behaviors, and the duration of emotions. Gender has been found to have an influence on the range of emotions experienced by individuals \citep{Hembree1988,Hyde1990,Frenzel2007}, and there are observed differences in emotional appraisals between genders \citep{pekrun2016academic}. Moreover, students' behaviors in CBLEs, such as engaging in on-task conversations or exhibiting off-task behavior, can redirect their learning trajectory \citep{Baker2004a,baker2010better}. Additionally, the duration of emotions has been noted as a contributing factor to the specific emotions experienced by students and the manner in which they experience them \citep{graesser2011theoretical,reis2018analysis,morais2023}.

Therefore, this RQ aims to determine whether the authors utilized information beyond the log data to enhance the automatic recognition of student emotions. The works of \cite{bosch2015accuracy,jensen2019generalizability,smeets2019affect,ocumpaugh2014population,cocea2010disengagement,yang2016exploring,wixon2014question,henderson2020improving,paquette2014sensor,hutt2019time,baker2012towards,baker2014extending,botelho2017improving} do not discuss the incorporation of new information in conjunction with emotions.

The papers of \cite{paquette2016sensor,kai2015comparison,pardos2014affective,jiang2018expert,defalco2018detecting} describe the collection of students' behavior alongside emotions. According to \cite{hintze2002best}, ``behavior is a result of the dynamic interplay between an individual and the environment," and it ``is thought to be specific to a situation." Specifically, students' behavior refers to their actions or attitudes related to the learning tasks, also known as learning-oriented behavior, which has been observed to impact learning outcomes \citep{perkins1965classroom}. For example, off-task behavior (related to not working on the content and not paying attention to the current learning task) has been found to negatively correlate with learning \citep{lee1999preliminary} and is associated with higher dropout rates \citep{finn1989withdrawing}. Conversely, on-task behavior has a positive impact on learning \citep{carroll1963model,bloom1976human}.

Only the papers of \cite{kai2015comparison} and \cite{jiang2018expert} reported the development of automatic behavior recognition. Similarly, \cite{tiam2019analysis} collected student actions by asking students to name their state, such as reading, thinking, writing, noting, and unfocused. However, none of these papers discuss the application of this information in the development of emotion detectors. 







\subsection{RQ2 - How are the sensor-free affect detectors being developed?}

The goal of this RQ is to find out the main methods, strategies, and approaches that the selected works used to make models that can detect students' emotions without sensors. This question was divided into the sub-questions below to deal with each aspect separately.


\subsubsection{RQ2.1 - Which features are mostly used to detect each emotion, and how are they captured and selected?}

The modeling process of sensor-free affect detectors involves the application of machine learning techniques to predict and classify emotions. This process is reliant on the extraction of input features derived from student actions within a CBLE, typically obtained from log data. In this study, a log is defined as a recorded entry that captures pertinent data pertaining to the actions undertaken by students within the CBLE. These logs are commonly stored in the CBLE database or specific log files.

Each log can contain multiple pieces of information, with each piece representing a unique ``feature" associated with the log. During the modeling process, feature engineering techniques are employed to generate additional features that encompass a wide range of characteristics and patterns found within the log data. This feature engineering step plays a crucial role in improving the accuracy of emotion detection and enables a more comprehensive understanding of student behavior \citep{botelho2019machine}. Notably, there are various approaches that can be utilized in feature engineering, including the extraction of temporal patterns to analyze the evolution of emotions over time. Additionally, the incorporation of contextual information or the consideration of individual characteristics allows for the accounting of situational and personal factors that influence emotions.

This RQ aims to investigate the most frequently utilized features for detecting each emotion and to explore the methods employed for collecting these features within CBLEs. Various approaches to feature selection are employed in works on sensor-free affect detection, driven by the diverse nature of emotions and their complex manifestation in educational settings. These approaches aim to identify features that are highly informative and can effectively capture the underlying emotional states of students. By selecting the most relevant features, researchers can enhance the performance and interpretability of their emotion detection models.

The choice of features depends on multiple factors, including the specific research objectives, available data, and characteristics of the CBLE. Researchers may consider features related to student behavior, such as interaction patterns, engagement levels, or temporal dynamics. Additionally, they may explore contextual information or demographic factors to enrich the feature set further. The ultimate goal is to capture a comprehensive representation of students' emotional experiences and tailor the detection models accordingly.

Through an investigation of the most frequently utilized features for detecting each emotion and an exploration of the methods employed for their collection, valuable insights can be gained into the specific indicators and patterns associated with different emotional states. This knowledge contributes to the development of robust and accurate emotion detection models within the context of computer-based learning environments.

The works used different features depending on the type of CBLE (such as MOOCs and intelligent tutoring systems) and the subject taught (such as math and physics). For example,  \cite{tiam2019analysis} used data from a CBLE that helps students with programming tasks. The authors made sensor-free affect detectors by considering all of the changes made to the code, such as insertions, deletions, compilations, and submissions of the assignments. \cite{yang2016exploring} collected the data from a MOOC. Thus, they focused on collecting features about posts and the student's actions in the environment through their computer mouse (clickstream).  In their study, \cite{jensen2019generalizability} and \cite{hutt2019time} collected data from Algebra Nation, a CBLE dedicated to math education. They gathered information on 22 specific features that were independent of specific content, such as video selection or quiz questions. Data analysis involved counting the occurrences of each feature within 30-second intervals and summing these counts across 5-minute window periods, with a maximum of 10 recorded activities for each 30-second chunk.  \cite{cocea2010disengagement} described the collection of data from the HTML-Tutor, which assists students in learning HTML. The authors developed emotion detectors based on features about the number of pages visited, the number of tests taken, the average time spent on pages, the average time spent on tests, the number of correctly answered tests, and the number of incorrectly answered tests.

In addition to the features that are unique to each type of CBLE and content, other works used more general features to build their sensor-free affect detectors. For example, some papers \citep{bosch2015accuracy,paquette2016sensor,kai2015comparison,smeets2019affect,pardos2014affective} described the use of features related to the number, type, and time of actions in the system, mouse clicks, keystrokes, and the aggregation of actions (e.g., the number of clicks in the last 5 seconds). Besides using general features, some works \citep{ocumpaugh2014population,defalco2018detecting,henderson2020improving,paquette2014sensor,baker2012towards,baker2014extending,botelho2017improving} added temporal and skill-based features and features based on the number of errors, correct answers, and hints requested. Still, some works describe the division of the features into groups \citep{jiang2018expert}, such as basic features (related to usage patterns), sequence features (related to sequential actions in the system), and threshold features (related to the amount of each feature when considering multiple students). Of all the works from this analysis, only \cite{wixon2014question} does not present information about the features considered during the model's development.








Several different features may be used to represent a single student's action with the system. However, using too many features during training can negatively impact the machine learning models. Some reasons are issues that make the results not generalizable for unseen data, such as noisy data and high sample dimensions. Feature selection is a common way to deal with a large number of features. Its goal is to find the features that best describe the data based on the target label, in this case, the student's emotions. We could identify four feature selection approaches used by the selected works. 
\begin{itemize}
    \item \textbf{Forward Selection:} This was the most cited approach from the selected papers \citep{paquette2016sensor,pardos2014affective,jiang2018expert,defalco2018detecting,henderson2020improving,paquette2014sensor,baker2012towards,baker2014extending}. This type of selection starts with an empty set of features and adds the feature that achieves the best performance each time. The selection stops when the performance does not increase, or the set achieves its maximum number of features.
    \item \textbf{Tolerance Analysis:} This was the second most frequently cited method \citep{bosch2015accuracy,jiang2018expert,tiam2019analysis}. This type of selection evaluates features' multicollinearity and eliminates highly collinear ones.  
    \item \textbf{Correlation-based:} This type of selection was used only by \cite{kai2015comparison}. The goal is to eliminate features with very little correlation with the target label. This approach also aims to remove features that are highly correlated with each other.
    \item \textbf{Backward Elimination:} This type of selection was used only by \cite{paquette2014sensor}. This approach can be seen as the opposite of the forward selection. It begins with a list of all features and removes those that have no significant effect on performance. It stops choosing when either the performance does not improve or when the set has a certain number of features.
\end{itemize}

In addition to reporting the approach used for feature selection, certain articles also provided details on additional aspects, including algorithms, thresholds, and validation strategies. Some papers even employed multiple strategies to enhance their affect detection models \citep{jiang2018expert,paquette2014sensor}. However, it is worth noting that certain works \citep{jensen2019generalizability,smeets2019affect,ocumpaugh2014population,cocea2010disengagement,wixon2014question,hutt2019time,botelho2017improving} did not perform or document any well-known feature selection processes, such as backward elimination. Nevertheless, some of these works utilized specific preprocessing techniques, such as summing up feature counts \citep{jensen2019generalizability,hutt2019time}.

From the papers that reported a feature selection phase, \cite{kai2015comparison,smeets2019affect,pardos2014affective,yang2016exploring,jiang2018expert,paquette2014sensor,hutt2019time,tiam2019analysis,baker2012towards,baker2014extending} described specific details on the selected features for each emotion model\footnote{The complete list of features related to the detection of each emotion can be found in this link: \url{https://bit.ly/3VTOmaC}.}. We identified common features selected independently of the emotion in most detectors. These features have to do with the number and frequency of different actions and tasks, the time it takes the student to do different actions, the history of help requests and correct or incorrect answers given by the student, and the number of actions done in different time windows. 

We were also able to find some features that were chosen for a specific emotion. Models were used to detect if someone was bored by looking at the number of wrong answers in a row, how fast they happened, and the chances of getting the answer right or guessing. Confusion was linked to characteristics of task types, task difficulty, the number of consecutive incorrect actions and their speed, the number of hints requested, the number of incomplete actions, the likelihood of answering or guessing correctly, and click patterns. Delight was related to features about the number of gamification trophies, the number of completed tasks or correct answers, and the time spent on different actions. Engagement was based on the number of completed tasks or correct answers, the history of actions, and the number of hints asked for. Frustration models used features about the number of hints requested, the number of consecutive incorrect actions and their speed, and the time of inactivity.

We also looked into how researchers synchronized the action logs and the emotion labels. Only some works  \citep{bosch2015accuracy,paquette2016sensor,jensen2019generalizability,pardos2014affective,cocea2010disengagement,paquette2014sensor,hutt2019time} have described this process. First, we have identified the grain size of the emotion labels, i.e., the assumed duration of an emotion. This duration has to do with the protocol the authors used to collect and annotate the student's emotions. The works by \cite{bosch2015accuracy,pardos2014affective,paquette2014sensor} collected emotions in a 20-second window. Other works \citep{jensen2019generalizability,cocea2010disengagement,hutt2019time} reported the emotion collection in windows of 1, 3, 5, and 10 minutes. This information is important because the synchronization between logs and emotions depends on this time. For example, suppose the work adopts a window size of 20 seconds, and a student performs 15 actions during this period. In that case, the authors must decide which approach to take for combining this information. We have identified two different approaches used by the selected papers. The first approach, used by \cite{paquette2016sensor,pardos2014affective}, replicates the emotion in these 15 action logs. Thus, each log inside the 20-second window will have the same synchronized emotion. The second approach, used by \cite{bosch2015accuracy,pardos2014affective}, aggregates the logs through some computation. For instance, the number of clicks, the average time spent on each action, and the number of hints asked for. So, the 15 logs will be combined into a single log with their information and a label for the emotion.

Another investigation we have performed is about the missing values. In a log, each feature represents different information. Some actions may not have enough information to fill in all the features. \cite{bosch2015accuracy,kai2015comparison} described the presence of missing values and the application of three different approaches to dealing with them. The first approach is zero imputation, used by \cite{bosch2015accuracy,kai2015comparison}, in which every single missing value is set to zero. The second approach is average imputation, used by \cite{kai2015comparison}, which takes the non-missing values of the same feature, computes their average, and sets the missing values to the computed average. The third approach, used by \cite{kai2015comparison}, is single imputation, in which the authors have built a tree-based decision model (M5) to predict the best value for the missing values of each feature.

\subsubsection{RQ2.2 - What are the main algorithms being used for sensor-free detectors?}
\label{sec:rq2.2}

\begin{figure}
    \centering
    \includegraphics[width=0.8\textwidth]{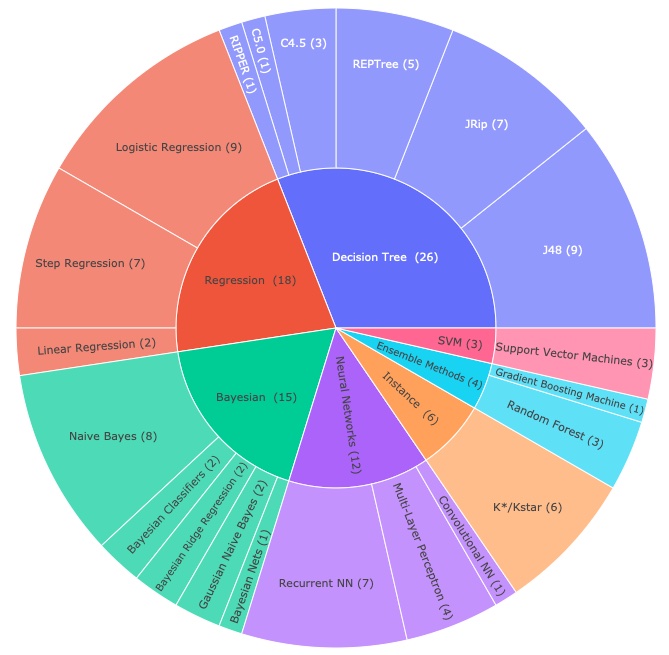}
    \caption{Distribution of articles by algorithm type. Specific algorithms were categorized into groups, including Decision Tree, Regression, Bayesian, Support Vector Machines, Instance-based Methods, Ensemble Methods, and Neural Networks. It is important to note that some algorithms could be classified under multiple types. A notable example is the Random Forest algorithm, an ensemble method that utilizes decision trees. In such cases, we chose to classify the algorithm based on its predominant methodology.}
    \label{fig:algorithms}
\end{figure}

We have also investigated the algorithms employed in the selected works, analyzing their development and analysis processes. While the number of tested algorithms varies, all the chosen works explicitly cite the algorithms they used. Figure \ref{fig:algorithms} presents the distribution of papers that utilized each algorithm. It's worth noting that the total count exceeds the number of considered papers, as some articles employed multiple algorithms simultaneously. 

We could observe that the common approach to developing sensor-free affect detectors is by starting with a set of algorithms, then selecting one model according to some goal. Even though some works did not provide any explanation for the algorithms' selection \citep{yang2016exploring,wixon2014question,hutt2019time,tiam2019analysis,bosch2015accuracy}, most papers \citep{kai2015comparison,pardos2014affective,ocumpaugh2014population,jiang2018expert,baker2012towards,baker2014extending} reported that they chose a specific set of common or standard classification algorithms that had previously been shown to be successful in building emotion detectors. Another reason is that the chosen algorithms can show different patterns in the data, but they are fairly conservative and less likely to overfit.  \cite{botelho2017improving} described the selection of three common recurrent network variants. Furthermore, some works selected their algorithms based on previous research, such as \citep{paquette2016sensor,smeets2019affect,defalco2018detecting,paquette2014sensor} citing \citep{baker2012towards}, \citep{paquette2016sensor,defalco2018detecting,paquette2014sensor} citing \citep{pardos2014affective}, \citep{henderson2020improving} citing \citep{defalco2018detecting}, \citep{jensen2019generalizability} citing \citep{hutt2019time}, and \citep{cocea2010disengagement} citing \citep{witten2005practical,mitchell1997machine}.

We also investigated the strategy they used to select the models.
We have identified three distinct strategies. The first and most commonly used strategy \citep{bosch2015accuracy,paquette2016sensor,kai2015comparison,smeets2019affect,jiang2018expert,henderson2020improving,paquette2014sensor,tiam2019analysis,baker2012towards,baker2014extending,botelho2017improving} is to select the model that achieves the best performance for each emotion. In this scenario, if the paper considers four different emotions, the authors choose the best model for each emotion. 
The second strategy selects the models according to a sample of the data. For instance, \cite{ocumpaugh2014population} selected the best model according to the data from different populations. 
Finally, the third strategy selects the model based on its successful application in previously developed work \citep{jensen2019generalizability,cocea2010disengagement}. In this case, the authors do not test any model. Instead, they choose the models according to the reported performance from previous work. This strategy differs from the first one in that it relies on the results of a model applied in a previous publication, rather than executing the model for the current work. It is worth noting that if the previously published work is by the same authors, the distinction between the first and third strategies may not be significant. However, we believe it is important to differentiate these strategies for the sake of clarity and thoroughness in our analysis.

The majority of papers developed single-class (or binary) classification models \citep{bosch2015accuracy,paquette2016sensor,kai2015comparison,smeets2019affect,pardos2014affective,ocumpaugh2014population,yang2016exploring,jiang2018expert,defalco2018detecting,henderson2020improving,paquette2014sensor,tiam2019analysis,baker2012towards,baker2014extending,botelho2017improving}. In this case, the authors have trained the classification models to learn the presence or absence of a single emotion, for example, to identify whether the student is confused or not. Some works have followed a multi-classification approach, in which the models are trained to differentiate between multiple classes \citep{botelho2017improving,cocea2010disengagement}. The model outputs whether the student is engaged, confused, neutral, and so on. \cite{jensen2019generalizability,wixon2014question} trained the models to infer the valence of the student's emotions, i.e., report whether the student is experiencing a positive or negative emotion. Finally,  \cite{wixon2014question,hutt2019time} inferred the intensity of each emotion or the level of presence or absence of an emotion, a regression problem.


During our investigation of the selected works, we examined the strategies employed for hyperparameter tuning and algorithm development in order to enhance the performance of sensor-free affect detectors. Hyperparameter optimization is a key focus in this process, aiming to identify the optimal parameter settings for a learning algorithm and thereby improving the model's overall performance. The incorporation of hyperparameter tuning and algorithm development within a systematic literature review on sensor-free affect detection is crucial, as it ensures that the advancements achieved through these techniques are thoroughly documented and analyzed. By identifying the specific approaches employed in the reviewed works, researchers and practitioners can gain valuable insights into the most effective strategies for optimizing affect detection models and driving advancements in the field.

Among the selected papers, \cite{smeets2019affect,jiang2018expert,hutt2019time,botelho2017improving} adopted hyperparameter optimization or fine-tuning techniques to optimize their models. Both \cite{smeets2019affect} and \cite{hutt2019time} employed the grid search strategy with cross-validation. This approach involves systematically exploring a predefined set of parameters for each algorithm, training and evaluating multiple models with different parameter combinations. Ultimately, the configuration that yields the best performance is selected as the final model.

In addition to grid search, \cite{jiang2018expert,hutt2019time,botelho2017improving} utilized various techniques to refine the parameters of their neural network models. These techniques included genetic algorithms, dropout regularization, and adjustments to the network topology. By employing these strategies, the authors achieved further optimization of the model's parameters, resulting in improved accuracy in detecting and classifying emotions.

\subsubsection{RQ2.3 - What measurements are used to identify the quality of the models (evaluation metrics)?}
\label{sec:rq2.3}

To evaluate and report on the performance of sensor-free affect detectors, researchers utilize various evaluation metrics depending on the type of problem, such as classification or regression. The selected papers employ different metrics, and their definitions are outlined below:

\begin{itemize}
\item \textbf{Cohen's Kappa} \citep{cohen1960coefficient}, also referred to as Kappa or K, measures the extent to which a detector outperforms random selection for binary classification tasks. A Kappa value of 0 indicates that the detector performs no better than random selection, while a value of 1 indicates perfect agreement with the annotated label. Cohen's Kappa is commonly used in classification problems.

\item \textbf{Fleiss Kappa} is a measure used to assess the agreement among two or more raters in categorical ratings. This metric is commonly employed in multi-label classification tasks. Similar to Cohen's Kappa, a value of 0 indicates that the group's choice was random, while a value of 1 signifies perfect agreement.

\item \textbf{Accuracy} measures the fraction of predictions that a model correctly classifies emotions. It can be employed in both binary and multi-class classification tasks. It is important to note that accuracy is most effective for evaluating classification problems with balanced classes, where the number of samples in each class is roughly equal.

\item \textbf{False Positive Rate} (FP) represents the ratio between the number of negative classes falsely classified as positive and the total number of actual negative classes.

\item \textbf{True Positive Rate} (TP), also known as ``\textbf{recall}," measures the ability of a model to correctly identify all instances of the positive class in a classification task. It quantifies the proportion of true positive predictions out of the total actual positive instances.

\item \textbf{Precision} measures the frequency with which a model correctly identifies the positive class in a classification task.  It quantifies the proportion of true positive predictions (correctly identified positive instances) out of all positive predictions made by the model. 

\item The \textbf{F1-score} is a metric also employed for evaluating binary classification models. Calculated as the harmonic mean of precision and recall, it can vary from 0, denoting the poorest performance, to 1, indicative of flawless precision and recall. The formula for computing the F1-score is as follows:

F1-score $= 2 \times \frac{{precision \times recall}}{{precision + recall}}$

\item \textbf{AUC-ROC}, also known as \textbf{AUC}, represents the ``Area Under the ROC Curve." The ROC Curve is a graphical representation that effectively illustrates the performance of a binary classifier by plotting the true positive rate against the false positive rate. AUC quantifies the probability of accurately identifying the presence or absence of an emotion, such as confusion or non-confusion, in binary classification tasks. A value of 0.5 indicates performance equivalent to random chance, while a value of 1 represents perfect performance. AUC-ROC is a widely used metric in classification problems.

\item \textbf{Spearman Correlation}, also known as $R_s$, is a correlation coefficient used to measure the strength and direction of the relationship between two non-parametric variables. It summarizes the monotonic relationship between the variables\footnote{In a monotonic relationship between two variables, the variables either consistently increase or decrease together, or one variable increases as the other decreases. Unlike in a linear relationship, the consistency in directionality of a monotonic relationship does not presuppose a constant rate of change. Conversely, a linear relationship indicates a fixed proportion of increase or decrease in one variable corresponding to each unit increase in the other variable, resulting in a straight line when represented graphically. Therefore, while every linear relationship qualifies as monotonic, the reverse is not necessarily true, given that not all monotonic relationships are linear.}, with values ranging from $-1$ to $+1$. A value of $-1$ indicates a perfect negative correlation, 0 indicates no correlation, and $+1$ represents a perfect positive correlation. Spearman Correlation is typically employed in regression problems.

\item \textbf{Pearson R} is a correlation coefficient that measures the strength and direction of the linear relationship between two continuous variables. Similar to Spearman Correlation, it is employed in regression problems, but evaluates linear relationships rather than monotonic relationships.

\item \textbf{Mean Absolute Error} (MAE) is the mean absolute difference between the predictions from a model and the actual emotion labels. It is primarily used in regression problems.

\end{itemize}

Table \ref{tab:evaluation_metrics} provides information about the number of works that have employed each metric and identifies the specific works in which they were employed.

\begin{table}[h]
\centering
\caption{Evaluation Metrics in Sensor-Free Affect Detection}
\label{tab:evaluation_metrics}
\begin{tabular}{|l|l|c|}
\hline
\textbf{Evaluation Metric} & \textbf{Papers occurrences} & \textbf{\#} \\ \hline
Cohen's Kappa & \citep{paquette2016sensor}, \citep{smeets2019affect} & 13 \\
 & \citep{pardos2014affective}, \citep{ocumpaugh2014population} & \\
 & \citep{yang2016exploring}, \citep{jiang2018expert} & \\
 & \citep{defalco2018detecting}, \citep{wixon2014question} & \\
 & \citep{henderson2020improving}, \citep{paquette2014sensor} & \\
 & \citep{baker2012towards}, \citep{baker2014extending} & \\
 & \citep{botelho2017improving} & \\ \hline
AUC & \citep{bosch2015accuracy}, \citep{paquette2016sensor} & 12 \\
 & \citep{kai2015comparison}, \citep{smeets2019affect} & \\
 & \citep{pardos2014affective}, \citep{ocumpaugh2014population} & \\
 & \citep{jiang2018expert}, \citep{defalco2018detecting} & \\
 & \citep{paquette2014sensor}, \citep{baker2012towards} & \\
 & \citep{baker2014extending}, \citep{botelho2017improving} & \\ \hline
Accuracy & \citep{cocea2010disengagement}, \citep{yang2016exploring} & 4 \\
 & \citep{henderson2020improving}, \citep{tiam2019analysis} & \\ \hline
F1-score & \citep{pardos2014affective}, \citep{henderson2020improving} & 2 \\ \hline
Spearman Correlation & \citep{jensen2019generalizability}, \citep{hutt2019time} & 2 \\ \hline
Fleiss Kappa & \citep{botelho2017improving} & 1 \\ \hline
TP and FP Rates & \citep{cocea2010disengagement} & 1 \\ \hline
Precision &  \citep{cocea2010disengagement} & 1 \\ \hline
Mean Absolute Error & \citep{cocea2010disengagement} & 1 \\ \hline
Person R & \citep{wixon2014question} & 1 \\ \hline
\end{tabular}
\end{table}

In general, we have observed that papers adopting the classification approach tend to employ Cohen's Kappa and/or AUC as their primary evaluation metrics. These metrics are preferred as they yield more reliable results, particularly when dealing with unevenly distributed data across different classes. On the other hand, papers focusing on the intensity of emotions captured by instruments, which entails a regression problem, utilize correlation metrics to compare the inferred intensity with the intensity captured by the instruments, often measured using Likert scales.

\subsubsection{RQ2.4 - How were the models trained and developed?}



We found that most works used the K-fold cross-validation strategy to train and validate the models. Cross-validation is a resampling method that splits the data into training and testing subsets. This strategy aims to divide the entire database into $k$ mutually exclusive training and testing sets. Detectors are trained and tested $k$ rounds. In each round, data from $k-1$ groups is used to train the detectors, and data from the remaining group to test the detector. Based on selected works, the value for $k$ varies between 10 \citep{paquette2016sensor,jensen2019generalizability,kai2015comparison,yang2016exploring,jiang2018expert,defalco2018detecting,hutt2019time,tiam2019analysis}, 6 \citep{baker2012towards}, 5 \citep{smeets2019affect,pardos2014affective,ocumpaugh2014population,wixon2014question,baker2014extending,botelho2017improving}, and 4 \citep{henderson2020improving}. A different approach was applied in \citep{bosch2015accuracy,paquette2014sensor}, where the authors used leave-one-out cross-validation. This method is similar to the k-fold cross-validation, the difference being the $k$ the number of students in the sample. Only \cite{cocea2010disengagement} did not report the validation process.

Some papers provided further details about the development of their machine learning pipeline, pointing out that the pre-processing steps, such as data normalization, class imbalance resampling, feature selection, and model training, were performed within each cross-validation fold. This approach is important to avoid data leakage between training and testing data. Also, another strategy being used by selected works is the use of a validation set. 
This approach splits the data into training, validation, and testing (\textit{hold-out}) subsets. The training and validation datasets, also called the development set, are used during the model's training. With cross-validation, these two datasets are automatically split according to the $k$ value. On the other hand, the hold-out dataset is not present at any point during model development, preventing data leakage during final performance analyses.

The validation level was not reported by \cite{cocea2010disengagement,yang2016exploring,tiam2019analysis}. All other works used student-level validation. In the student-level validation method, data from the same student are either in the training or testing set but never in both. This method was used to provide more confidence that the model generalizes well to new students. 

\subsubsection{RQ2.5 - What are the main tools/technologies being used to develop these models?}

RapidMiner \citep{paquette2016sensor, kai2015comparison, ocumpaugh2014population, jiang2018expert, defalco2018detecting, wixon2014question, baker2012towards} and WEKA \citep{bosch2015accuracy, cocea2010disengagement, tiam2019analysis} are the most commonly used tools. These tools offer a range of machine learning functionalities for data science and data mining tasks. RapidMiner\footnote{\url{https://rapidminer.com}} provides a user-friendly graphical interface that enables the development of machine learning solutions through a drag-and-drop approach, reducing development time and potential errors. Similarly, WEKA\footnote{\url{https://www.cs.waikato.ac.nz/ml/weka}} offers a collection of machine learning algorithms and tools for data preparation, classification, regression, clustering, association, and visualization. The ease of use and versatility of these tools make them popular choices for prototyping and testing sensor-free affect detection models.

In addition to RapidMiner and WEKA, selected works have utilized various Python libraries such as Theano \citep{botelho2017improving}, Lasagne \citep{botelho2017improving}, scikit-learn \citep{jensen2019generalizability, hutt2019time}, TensorFlow \citep{hutt2019time}, and Keras \citep{hutt2019time}. Theano\footnote{\url{https://pypi.org/project/Theano}} is a library that facilitates the definition, optimization, and evaluation of mathematical expressions involving multi-dimensional arrays. Lasagne\footnote{\url{https://lasagne.readthedocs.io}} is a lightweight library built on top of Theano, designed specifically for building and training neural networks. scikit-learn\footnote{\url{https://scikit-learn.org}} integrates seamlessly with other Python scientific libraries and offers a wide range of state-of-the-art algorithms for machine learning tasks. TensorFlow\footnote{\url{https://www.tensorflow.org}} is an open-source library that simplifies the development of neural networks and numerical computations. Keras\footnote{\url{https://keras.io}} is a user-friendly deep learning API that runs on top of TensorFlow, enabling rapid experimentation with machine learning models.

It is worth mentioning that some studies \citep{smeets2019affect, pardos2014affective, yang2016exploring, henderson2020improving, paquette2014sensor} did not explicitly describe the tools or technologies used in their research.

Over time, the technology landscape for developing machine learning models, including sensor-free affect detectors, has evolved. While RapidMiner and WEKA remain widely used tools, the emergence of Python libraries such as scikit-learn, TensorFlow, and Keras has introduced new possibilities and improved performance in this field. Python's versatility, combined with the popularity of deep learning frameworks like TensorFlow and Keras, has contributed to their increased adoption in recent years.

\subsubsection{RQ2.6 - How many affect emotion labels, action logs and features are being considered?}

By analyzing the selected works, there is no standard quantity of labels and features, and it varies for each work.

First, we investigated the number of logs collected and used by the selected works. Only two papers reported the number of logs examined: \citep{jiang2018expert} (146K) and \citep{cocea2010disengagement} (1K). On the other hand, only two papers did not present the number of features used to develop their detection models: \citep{wixon2014question,tiam2019analysis}. All other selected works have reported the number of features considered. Again, this number is different for each work, ranging from 5 to 7 for \citep{baker2014extending}, from 20 to 40 for \citep{paquette2016sensor,jensen2019generalizability,cocea2010disengagement,defalco2018detecting,henderson2020improving,hutt2019time}, from 40 to 60 for \citep{smeets2019affect,yang2016exploring}, from 60 to 80 for \citep{kai2015comparison,ocumpaugh2014population}, from 110 to 130 for \citep{bosch2015accuracy,paquette2014sensor}, 172 for \citep{pardos2014affective}, 204 for \citep{botelho2017improving}, 232 for \citep{baker2012towards}, and 249 for \citep{jiang2018expert}. Some papers have reported a ``feature selection stage," in which a certain number of features that are not considered ideal for each model are eliminated. However, most papers do not describe whether there was a limitation on the maximum number of selected features or whether it was decided according to the feature selection algorithms. 

We also examined the distribution of the number of emotion labels used in the studies. This range varied from less than 1K \citep{baker2012towards} to between 1K and 2K \citep{bosch2015accuracy}, 2K and 3K \citep{kai2015comparison,baker2014extending}, 3K and 4K \citep{paquette2016sensor,pardos2014affective,defalco2018detecting,henderson2020improving}, 4K and 5K \citep{paquette2014sensor}, 5K and 6K \citep{yang2016exploring,jiang2018expert}, and 7K and 8K \citep{botelho2017improving}. Notably, one study reported an unusually high number of over 133K emotion labels \citep{hutt2019time}, which was identified as an outlier and subsequently removed from our analysis.

Excluding the outlier, the average number of collected emotion labels across the studies was approximately 3.5K. However, it should be noted that some studies utilized only a subset of the collected labels, with the number ranging from less than 1K \citep{paquette2016sensor,defalco2018detecting,henderson2020improving} to between 1K and 2K \citep{bosch2015accuracy,kai2015comparison,paquette2014sensor}. On average, the number of emotion labels used in the studies was approximately 2.3K. Notably, the papers by \cite{jensen2019generalizability, smeets2019affect, ocumpaugh2014population,cocea2010disengagement,wixon2014question,tiam2019analysis} did not provide explicit information regarding the number of emotion labels used in their respective studies.




In addition to the total number of emotion labels, we looked into how the works deal with unequal classes. This is an important subject to consider because, in classification, different numbers of instances representing each class, in this case, the student's emotions, can impose a load-balancing problem\footnote{More details about class imbalance problem in classification tasks can be found in \citep{ali2013classification}.}, leading to wrong results or interpretations of the results. Selected works used two resampling strategies to deal with imbalanced classes. The first strategy is known as ``oversampling'', applied by \cite{bosch2015accuracy,paquette2016sensor,kai2015comparison,jiang2018expert,defalco2018detecting,henderson2020improving,baker2014extending,botelho2017improving}, which creates samples of the minor representative emotions. They used two approaches to create the new samples. The first one is to clone the minority samples, applied by \cite{kai2015comparison,jiang2018expert,defalco2018detecting,baker2014extending,botelho2017improving}, and the second one is through data augmentation, using the SMOTE technique, applied by \cite{bosch2015accuracy,henderson2020improving}. Another strategy to deal with imbalanced data is downsampling or undersampling, applied by \cite{bosch2015accuracy,smeets2019affect,botelho2017improving}, which consists of reducing the sample from the majority class. As the authors of SMOTE \citep{chawla2002smote} suggested, the best way to use SMOTE would be to randomly undersample the class that makes up the majority. This approach was applied in one selected work \citep{bosch2015accuracy}.
 \cite{pardos2014affective,paquette2014sensor,baker2012towards} reported some resampling, but did not describe which they used. Besides,  \cite{jensen2019generalizability,ocumpaugh2014population,cocea2010disengagement,yang2016exploring,wixon2014question,hutt2019time,tiam2019analysis} did not present any information about resampling or data imbalance.

\subsubsection{RQ2.7 - What are the best results and algorithms for detecting each emotion?}

\begin{figure}[ht]%
    \centering
    \includegraphics[width=1\textwidth]{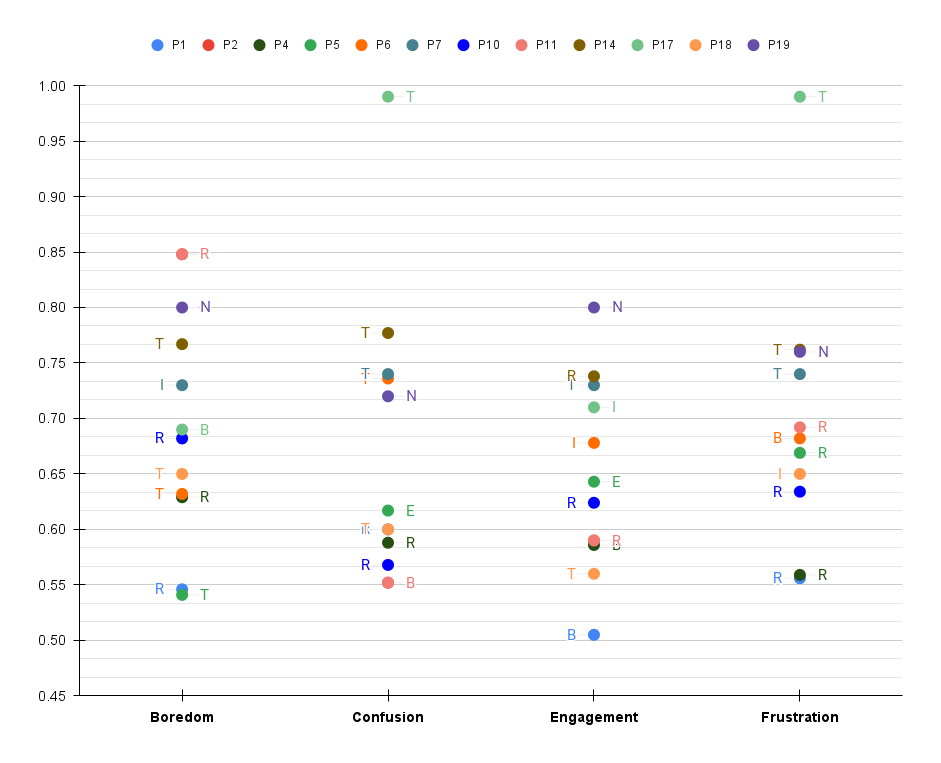}
    \caption{Scatter plot showing the best results for each study on sensor-free affect detectors, measured in AUC. Each point on the graph represents the optimal result found in a study (identified by their ID in Table \ref{tab:selected_works} in the legend). The labels next to the points represent the type of algorithm (according to Figure \ref{fig:algorithms}) that achieved the highest result, which are:  
    \textbf{R} = Regression
, \textbf{B} = Bayesian
, \textbf{T} = Decision Tree
, \textbf{E} = Ensemble-based
, \textbf{I} = Instance-based
, \textbf{N} = Neural Networks.}
    \label{fig:best_models}
\end{figure}

This RQ aims to present the best models reported in each selected work. To answer this question, we analyzed each work, searching for the best result in detecting each emotion. Some studies reported different results when considering different types of data altogether (multi-modal combining interaction data with video, text, sensors, etc.). However, we considered only the results based on interaction data. Because we have so many emotions (25 emotions according to RQ1.1) and so many performance metrics (11 metrics according to RQ2.4), we selected the most frequent emotions, reporting their results with the most commonly used metrics. So, for emotions, we looked at boredom, confusion, engagement, and frustration, with more than 16 occurrences each (see Section \ref{sec:rq1.1}). Figure \ref{fig:best_models} presents the AUC values for the detection of boredom, confusion, engagement, and frustration, and the algorithms that achieved each value. The color of each point identifies the results of one of the selected works in a given metric. Papers that did not report the AUC metrics for one of the four emotions were not considered.

It is important to notice that each work uses different information to figure out what emotions someone is feeling and gets its data from different CBLEs, contexts, and contents. Therefore, Figure \ref{fig:best_models} aims to provide an overview of the results presented in the sensor-free affect detection field. In addition, the list of algorithms is not extensive because it only shows the algorithms that achieved the best results. RQ2.2 shows the complete list of algorithms (see Section \ref{sec:rq2.2}).

\subsection{RQ3 - For which scenario are the sensor-free affect detectors being used?}

To address RQ3, we devised five sub-questions to delve deeper into the research sample, educational environment type, learning domain, and goals and purposes of the works. 

\subsubsection{RQ3.1 - What are the characteristics of the sample?}

This RQ identifies the sample in which the data is being collected. Most research has been conducted on regular K–12 education. The works have collected data from $6^{th}$ \citep{smeets2019affect,jiang2018expert}, $8^{th}$ \citep{pardos2014affective,paquette2014sensor}, $8^{th}$  to $9^{th}$ \citep{bosch2015accuracy,kai2015comparison}, $7^{th}$ to $10^{th}$ \citep{wixon2014question}, and $6^{th}$ to $12^{th}$ grades \citep{jensen2019generalizability}. Works also collected data from students in the military academy \citep{paquette2016sensor,defalco2018detecting,henderson2020improving} and $1^{st}$ year programming classes \citep{tiam2019analysis}. The papers by \cite{ocumpaugh2014population,cocea2010disengagement,yang2016exploring,hutt2019time,baker2012towards,baker2014extending,botelho2017improving} did not provide information about the students' sample.

We also looked into the sample size or the number of students from whom the action logs were taken. We found out that papers collected data from the following number of students: less than 50 \citep{cocea2010disengagement}, between 73 and 93 \citep{jiang2018expert,tiam2019analysis,baker2012towards}, between 119 and 153 \citep{bosch2015accuracy,paquette2016sensor,kai2015comparison,defalco2018detecting,henderson2020improving,baker2014extending}, between 229 and 326 \citep{pardos2014affective,wixon2014question,paquette2014sensor}, 646 \citep{botelho2017improving}, 4281 \citep{yang2016exploring}, and 69174 students \citep{jensen2019generalizability,hutt2019time}. 

Only three papers reported the students' age, in which \citep{smeets2019affect} collected data from students between 4 and 12 years old and \citep{defalco2018detecting,henderson2020improving} from students between 18 and 22 years old. 

Most of the selected works collected data from students using the CBLE in the school's lab \citep{bosch2015accuracy,paquette2016sensor,jensen2019generalizability,kai2015comparison,pardos2014affective,ocumpaugh2014population,yang2016exploring,jiang2018expert,defalco2018detecting,wixon2014question,henderson2020improving,paquette2014sensor,hutt2019time,tiam2019analysis,baker2012towards,baker2014extending,botelho2017improving}. \citep{smeets2019affect,yang2016exploring} reported collecting action logs from students using the CBLE at home, whereas \cite{cocea2010disengagement} did not specify where the students were while using the CBLE. 

Of the works that collected data from schools, only \cite{bosch2015accuracy,kai2015comparison,jiang2018expert} described students as attending a public school. \cite{smeets2019affect} also described that students paid for a subscription to use the CBLE. Other selected works did not contain this information.

Some studies have reported additional information about the sample, such as genders \citep{bosch2015accuracy,paquette2016sensor,kai2015comparison,defalco2018detecting,henderson2020improving}, ethnicities \citep{pardos2014affective}, schools \citep{jensen2019generalizability,paquette2014sensor,hutt2019time}, and countries \citep{tiam2019analysis}.

We also analyzed the period over which the selected works collected student log data. The results show that it varies from a few days (1 to 7 days) \citep{bosch2015accuracy,paquette2016sensor,kai2015comparison,cocea2010disengagement,jiang2018expert,wixon2014question,baker2012towards,baker2014extending}, weeks \citep{yang2016exploring}, and a whole school year \citep{jensen2019generalizability,ocumpaugh2014population,hutt2019time,botelho2017improving,pardos2014affective}. Regarding the duration of the sessions, they vary between 30 and 45 minutes \citep{jiang2018expert,tiam2019analysis}, 55 minutes \citep{bosch2015accuracy,kai2015comparison}, and one to two hours \citep{paquette2016sensor,defalco2018detecting,henderson2020improving,pardos2014affective}.  \cite{smeets2019affect,paquette2014sensor} did not provide this information.  

We also identified where the sample came from. Most of the sensor-free affect detection research collected data from students in the United States \citep{bosch2015accuracy,paquette2016sensor,jensen2019generalizability,kai2015comparison,pardos2014affective,ocumpaugh2014population,yang2016exploring,jiang2018expert,defalco2018detecting,wixon2014question,henderson2020improving,paquette2014sensor,hutt2019time,baker2012towards,baker2014extending,botelho2017improving}. A few studies were done in other countries, like the Netherlands \citep{smeets2019affect}, Japan \citep{tiam2019analysis}, the Philippines \citep{tiam2019analysis}, and Germany \citep{cocea2010disengagement}.

\subsubsection{RQ3.2 - Which type of educational environment?}

The selected works prominently feature CBLEs that can be categorized into several types, including Intelligent Tutoring Systems (ITS), game-based environments, web-based platforms, Massive Open Online Courses (MOOCs), coding IDEs, and virtual worlds. Table \ref{tab:cbles} provides a categorization summary for each selected work.

In the ITS category, we encountered a diverse range of systems, such as Inq-ITS \citep{paquette2014sensor}, HTML-Tutor \citep{cocea2010disengagement}, ASSISTments \citep{pardos2014affective,ocumpaugh2014population,botelho2017improving}, Cognitive Tutor Algebra I \citep{baker2012towards}, Betty’s Brain \citep{jiang2018expert}, and Wayang Outpost \citep{wixon2014question}.

Transitioning our attention to game-based environments, we identified various platforms like Physics Playground \citep{bosch2015accuracy,kai2015comparison}, vMedic - TC3Sim \citep{paquette2016sensor,defalco2018detecting,henderson2020improving}, and Squla \citep{smeets2019affect}.

Our research also uncovered unique CBLEs that were the subject of individual studies. These include the virtual world EcoMUVE \citep{baker2014extending}, a custom Integrated Development Environment (IDE) for coding \citep{tiam2019analysis}, and a study conducted on the MOOC platform, Coursera \citep{yang2016exploring}.

In addition, we came across two studies that utilized data from Algebra Nation \citep{jensen2019generalizability,hutt2019time}. Algebra Nation is a web-based platform specifically designed to assist in mathematics learning, with a focus on algebra. It combines the use of instructional videos, practice exercises, and group discussions to enrich the learning experience.

\begin{table}[htbp]
\centering
\caption{Computer-Based Learning Environments (CBLE) categories and platforms}
\label{tab:cbles}
\begin{tabular}{|c|l|}
\hline
\textbf{Category} & \textbf{Platforms} \\ \hline
ITS & Inq-ITS \citep{paquette2014sensor} \\
    & HTML-Tutor \citep{cocea2010disengagement} \\
    & ASSISTment \citep{pardos2014affective,ocumpaugh2014population} \\
    & \hspace{65pt} \citep{botelho2017improving} \\
    & Cognitive Tutor Algebra I \citep{baker2012towards} \\
    & Betty’s Brain \citep{jiang2018expert} \\
    & Wayang Outpost \citep{wixon2014question} \\ \hline
Game-based & Physics Playground \citep{bosch2015accuracy,kai2015comparison} \\
    & vMedic - TC3Sim \citep{paquette2016sensor} \\
    & \hspace{88pt} \citep{defalco2018detecting} \\
    & \hspace{88pt} \citep{henderson2020improving} \\
    & Squla \citep{smeets2019affect} \\ \hline
MOOCs & Algebra Nation \citep{jensen2019generalizability,hutt2019time} \\ \hline
Individual Studies & EcoMUVE \citep{baker2014extending} \\
                    & Coursera \citep{yang2016exploring} \\
                    & Custom IDE for coding \citep{tiam2019analysis} \\ \hline
\end{tabular}
\end{table}

\subsubsection{RQ3.3 - In which learning domain (content)?}



This RQ investigates the learning domain in which the selected works apply or develop their sensor-free affect detectors. We found out that math is the most studied subject, with nine works \citep{jensen2019generalizability,smeets2019affect,pardos2014affective,ocumpaugh2014population,yang2016exploring,wixon2014question,hutt2019time,baker2012towards,botelho2017improving}. The second and third most studied subjects are medicine \citep{paquette2016sensor,defalco2018detecting,henderson2020improving} and science \citep{jiang2018expert,paquette2014sensor,baker2014extending}, with three works each. The fourth and fifth most studied subjects are programming \citep{cocea2010disengagement,tiam2019analysis} and physics \citep{bosch2015accuracy,kai2015comparison}, with two works each. The final subject on the list is economics, which is taught together with mathematics in work described in the paper \citep{yang2016exploring}.

\subsubsection{RQ3.4 - What are the goals and purposes of the works? }

Regarding the \textbf{goal of the paper}, we refer to what the authors intended to achieve with their work. The goals usually set the focus of the research, being the targets that the authors aim to hit through their experiments, analyses, developments, and evaluations. In this analysis, we categorized the works into five distinct groups, which are described below. Table \ref{tab:goal} presents the works that fall into each goal category.

\begin{itemize}
    \item \textbf{Building emotion detectors for different samples or data}: These papers aimed at developing sensor-free affect detectors by considering samples collected from different populations, cultures, educational levels, backgrounds, or environments. The goal was to achieve better generalization of the results.
    \item \textbf{Comparing different types of emotion detection}: These papers developed sensor-free affect detectors by considering different types of data, aiming to determine the most effective approach for each scenario.
    \item \textbf{Improving detection performance and coverage}: These papers focused on developing various strategies to enhance the performance of sensor-free affect detectors.
    \item \textbf{Studying the relationship between the detected emotions and other constructs}: Papers in this group aimed to build sensor-free affect detectors capable of inferring students' emotions and studying the relationship between the detected emotions and different constructs. For example, they may have explored predicting end-of-year learning outcomes or investigating the impact of self-reporting emotions on students' emotions.
    \item \textbf{Validating interventions}: These papers aimed to detect students' emotions through sensor-free detectors to validate the effectiveness of different learning interventions.
\end{itemize}

\begin{table}[htbp]
\centering
\caption{Categorization of works based on goal}
\label{tab:goal}
\begin{tabular}{|l|l|}
\hline
\textbf{Category} & \textbf{Works} \\ \hline
Build emotion detectors for different sample or data & \citep{jensen2019generalizability} \\
 & \citep{smeets2019affect} \\
 & \citep{ocumpaugh2014population} \\
 & \citep{cocea2010disengagement} \\
 & \citep{yang2016exploring} \\
 & \citep{hutt2019time} \\
 & \citep{baker2014extending} \\ \hline
Compare different types of emotion detection & \citep{paquette2016sensor} \\
 & \citep{kai2015comparison} \\
 & \citep{jiang2018expert} \\
 & \citep{defalco2018detecting} \\
 & \citep{henderson2020improving} \\
 & \citep{tiam2019analysis} \\ \hline
Improve detection performance and coverage & \citep{bosch2015accuracy} \\
 & \citep{paquette2014sensor} \\
 & \citep{baker2012towards} \\
 & \citep{botelho2017improving} \\ \hline
Study the relationship between the detected emotions and  & \citep{pardos2014affective} \\
other constructs  & \citep{wixon2014question} \\ \hline
Validation of interventions & \citep{yang2016exploring} \\
 & \citep{defalco2018detecting} \\ \hline
\end{tabular}
\end{table}


Regarding the \textbf{purpose} of the works, we refer to the practical application and functionality of the research results, or what is intended to be done with the detection model. Purposes are often related to the impact and applicability of the research results in the real world. Based on the description of the papers, we identified seven categories, listed below. Table \ref{tab:purpose} presents the works corresponding to each identified purpose category.

\begin{itemize}
    \item Providing meaningful and effective interventions and responding to emotions in real-time (emotion regulation).
    \item Designing and building intelligent learning interfaces.
    \item Studying and improving other constructs based on the detected emotions, such as behavior and learning outcomes, or exploring the emotions themselves through modeling.
    \item Enhancing performance and applicability for different students, types of data, and computer-based learning environments (CBLEs).
    \item Keeping track of student emotions.
    \item Conducting discovery with models.
    \item Offering an alternative to sensor-based emotion detectors.
\end{itemize}

\begin{table}[htbp]
\centering
\caption{Categorization of works based on purpose}
\label{tab:purpose}
\begin{tabular}{|l|l|}
\hline
\textbf{Category} & \textbf{Works} \\ \hline
Provide meaningful and effective interventions and  
& \citep{smeets2019affect} \\ 
respond to emotions in real-time
 & \citep{cocea2010disengagement} \\
 & \citep{jiang2018expert} \\
 & \citep{defalco2018detecting} \\
 & \citep{henderson2020improving} \\
 & \citep{paquette2014sensor} \\
 & \citep{hutt2019time} \\
 & \citep{tiam2019analysis} \\
 & \citep{baker2012towards} \\
 & \citep{baker2014extending} \\ \hline
Design/build intelligent learning interfaces & \citep{bosch2015accuracy} \\
 & \citep{henderson2020improving} \\
 & \citep{paquette2014sensor} \\
 & \citep{hutt2019time} \\
 & \citep{tiam2019analysis} \\ \hline
Study/improve other constructs based on the detected emotions & \citep{jensen2019generalizability} \\
 & \citep{pardos2014affective} \\
 & \citep{yang2016exploring} \\
 & \citep{wixon2014question} \\ \hline
Performance and applicability for different students and & \citep{kai2015comparison} \\  types of data and CBLEs &   \citep{ocumpaugh2014population} \\
 & \citep{botelho2017improving} \\ \hline
Keep track of student emotions & \citep{cocea2010disengagement} \\
 & \citep{yang2016exploring} \\ \hline
Discovery with models & \citep{baker2014extending} \\ \hline
Alternative for sensor-based emotion detectors & \citep{paquette2016sensor} \\ \hline
\end{tabular}
\end{table}

\subsection{RQ4 - What are the trends and main ideas for future research?}

The RQ4 aims to provide an overview of the ideas for future research pointed out by the selected works when considering the field of sensor-free affect detection. We broke this question down into specific sub-questions about how the developed models can be used in the real world and how they can be put into production.  We were also interested in identifying the trends and the main scientific stakeholders on this subject.

\subsubsection{RQ4.1 - What is the generalization performance of the developed detectors and how could they  be applied in production?}






This RQ verifies what the selected works described about the generalization performance of their sensor-free affect detection models for different learners' profiles. We also examined whether the selected works applied their models in production to detect students' emotions in real-time. We found that only seven works, or 37\%, mentioned how the models could be used to detect emotions for other learners' profiles. Two of these studies say that the results cannot be generalized because the sample size was too small  \citep{tiam2019analysis} or there were too few courses  \citep{yang2016exploring}.

Other works presented to what extent their results are generalizable. 
\cite{paquette2016sensor} described that their posture-based detectors might achieve a better generalization performance because they consider the learner behavior outside the software, making the detectors system-independent. 
\cite{ocumpaugh2014population} collected data from different populations, including those in cities, suburbs, and rural areas. They showed that the detection models could only be used for different populations if they were included in the sample. 
\cite{jensen2019generalizability} trained models over a different number of students and found that models trained with fewer than 1500 students did not generate stable scores or predictions. 
\cite{cocea2010disengagement} mentioned that the proposed approach is generalized to systems other than e-learning.  
\cite{hutt2019time} used trained models on data from different content, demonstrating that models trained for one content could be used on data for another.

Only one of the chosen works \citep{defalco2018detecting} talked about how the developed sensor-free detection models could be used in production.  Their models were integrated into a CBLE and used by the students in an experiment. This experiment aimed to respond to the students' frustration through feedback messages. The authors tried out three different kinds of motivational feedback to determine which helped students learn the most.

\subsubsection{RQ4.2 - What are the trends and future directions in sensor-free affect detection?}

This research question (RQ) seeks to identify the current trends and potential future directions in sensor-free affect detection, as outlined in the selected works. ``Trends'' and ``future works'' are two distinct aspects in scientific research. ``Trends'' refer to the current or emerging directions within a field, reflecting ongoing developments and noteworthy patterns. These can include novel findings, methodologies, technologies, or approaches that have gained prominence. On the other hand, ``future works'' refer to areas that require further investigation or exploration. They represent the gaps, unresolved issues, or unanswered questions within the existing body of knowledge. Future works are proposed as potential research endeavors based on the limitations of the current study, preliminary findings, or opportunities for expanding upon existing knowledge.

To identify the trends, we first analyzed the authors' suggestions when reporting their results. We then synthesized the main trends into six categories, listed in order of their frequency:

\begin{itemize}
\item \textbf{Generalization:} This trend focuses on collecting or testing the detection models across different populations, CBLEs, and contexts \citep{paquette2016sensor,ocumpaugh2014population,yang2016exploring,wixon2014question,botelho2017improving}.

\item \textbf{Multimodal:} This trend involves using multiple data sources (beyond just interaction logs) to enhance the accuracy of detecting students' emotions \citep{bosch2015accuracy,paquette2016sensor,kai2015comparison,henderson2020improving,tiam2019analysis}.

\item \textbf{Intervention:} This trend pertains to works aiming to provide meaningful and effective interventions based on the detected emotions, such as emotional feedback, content adaptation, teacher assistance, etc. \citep{smeets2019affect,pardos2014affective,defalco2018detecting,baker2014extending}.

\item \textbf{Knowledge Discovery:} This trend involves the discovery of new knowledge, constructs, and understanding of students' emotions based on the developed models \citep{jiang2018expert,hutt2019time,baker2012towards,baker2014extending}.

\item \textbf{Real-time Detection:} This trend is derived from works that describe the integration of the models into CBLEs for real-time detection and decision-making during student learning \citep{pardos2014affective,yang2016exploring,hutt2019time}.

\item \textbf{Performance Improvement:} This trend identifies works aiming to enhance the performance of detection models using different technologies or techniques \citep{jensen2019generalizability,paquette2014sensor}.
\end{itemize}

We also identified future works reported by the selected papers to provide an overview of potential directions in the field. The following future research suggestions were identified:

\begin{itemize}
\item Utilize multi-modal data sources \citep{bosch2015accuracy,kai2015comparison,henderson2020improving};
\item Implement real-time interventions and adaptive interfaces (redesign the CBLE based on detected student emotion) \citep{bosch2015accuracy,yang2016exploring,paquette2014sensor,tiam2019analysis,baker2014extending};
\item Enhance data mining performance \citep{paquette2016sensor,jensen2019generalizability,defalco2018detecting}, by applying:
\begin{itemize}
\item Ensemble approaches \citep{hutt2019time};
\item Deep learning \citep{botelho2017improving};
\item Feature construction/selection \citep{jensen2019generalizability,smeets2019affect,defalco2018detecting,hutt2019time,tiam2019analysis,baker2012towards};
\item Algorithm selection \citep{smeets2019affect};
\item Aggregation methods \citep{baker2012towards};
\item Improved sampling strategies \citep{henderson2020improving};
\end{itemize}
\item Understand how emotions influence learning and vice-versa:
\begin{itemize}
\item How emotions influence learning \citep{pardos2014affective};
\item How knowledge shapes emotions \citep{ocumpaugh2014population};
\item How behaviors correspond to emotions \citep{baker2012towards};
\item How CBLE aspects interfere with emotions \citep{baker2014extending};
\item How emotions predict educational outcomes \citep{hutt2019time};
\end{itemize}
\item Utilize student's affect and behavior data for professor recommendations and interventions \citep{pardos2014affective};
\item Validate on broader populations/generalizations, contexts, and CBLEs \citep{jensen2019generalizability,ocumpaugh2014population,jiang2018expert,wixon2014question,hutt2019time,baker2012towards, botelho2017improving};
\item Collect more samples of underrepresented emotions, such as confusion and frustration, in the classroom \citep{yang2016exploring,botelho2017improving}.
\end{itemize}


\subsubsection{RQ4.3 - What is the current status/overview of the research area (publication avenues, year, authors)?}

\begin{figure}
\centering
\includegraphics[width=1\textwidth]{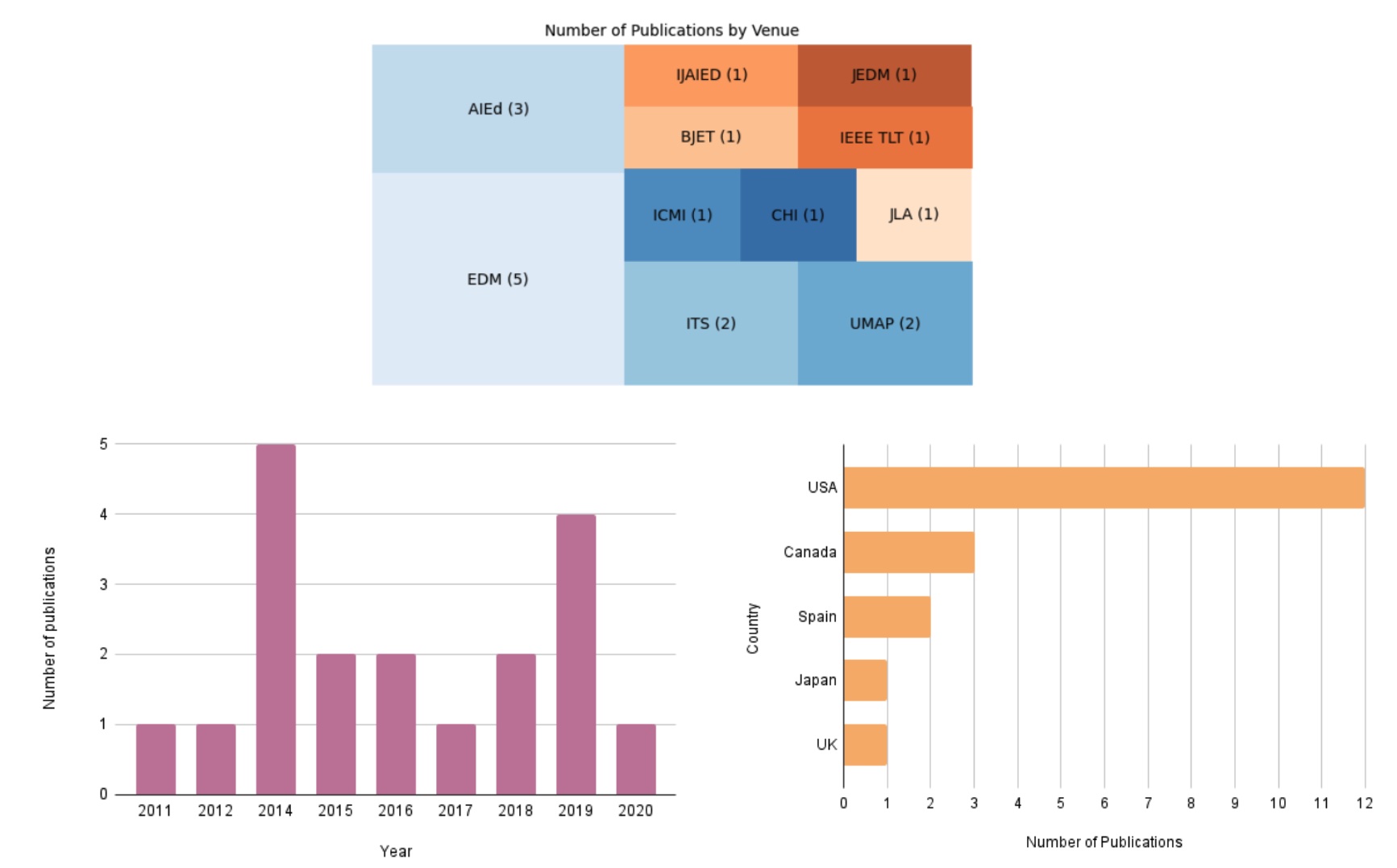}
\caption{Overview of publication venues, countries, and years}
\label{fig:publications_overview}
\end{figure}

This research question aims to provide an overview of the research area. We examined the journals and conference proceedings in which the selected papers were published, as well as the number of publications per year and country. Figure \ref{fig:publications_overview} illustrates these findings.

The conferences and journals frequently used by authors in this area include: International Conference on Educational Data Mining (EDM), International Conference on Artificial Intelligence in Education (AIEd), International Conference on Intelligent Tutoring Systems (ITS), International Conference on User Modeling, Adaptation, and Personalization (UMAP), International Conference on Multimodal Interaction (ICMI), Conference on Human Factors in Computing Systems (CHI), Journal of Learning Analytics (JLA), British Journal of Educational Technology (BJET), International Journal of Artificial Intelligence in Education (IJAIED), IEEE Transactions on Learning Technologies (IEEE TLT), and Journal of Educational Data Mining (JEDM). The frequency in each publication vehicle is illustrated Figure \ref{fig:publications_overview}.

We compiled the names of all authors from the selected works, totaling 55 researchers. The leading authors in this field are Ryan Baker (12 publications), Luc Paquette (6 publications), Jaclyn Ocumpaugh (5 publications), Sidney K. D'Mello (4 publications), and Sujith M. Gowda (4 publications). The remaining researchers have three or fewer publications on this subject.

\section{Main Findings}

This section synthesizes key findings from our Systematic Literature Review (SLR) on sensor-free affect detection.

Boredom, confusion, frustration, and engagement are the primary emotions detected in 84\% of the 19 selected studies. The frequency of these emotions reflects their common manifestation in students during complex or technology-assisted learning tasks. However, further research into other emotions contingent on the learning environment type is encouraged. For example, \cite{morais2023} reported frequent instances of surprise due to the limited feedback in their step-based tutoring system.

The reviewed studies adopted various methods to identify learners' emotions, including human observer annotations, student self-reports, crowdsourcing, and log file annotation. The Behavioral Observation of Students in Schools (BROMP) protocol is often employed for encoding students' emotions in real-time, mainly due to its creators' high citation rate. However, it precludes the acquisition of sequential emotional data due to the round-robin annotation approach. Self-reporting is the second most prevalent method, followed by crowdsourcing. Some studies used log file annotation, which relies solely on log data to annotate emotions.

When developing sensor-free emotion classifiers, it's crucial to consider the "grain size," or the time duration during which emotions are annotated, and actions are captured. As discussed in Section \ref{sec:req1.3}, this duration varies significantly across studies, affecting classifier performance and being constrained by the annotation method.

The Control-Value Theory (CVT) is the most frequently referenced psychological theory for defining emotions in the selected works. CVT centers on learning-oriented emotions and suggests emotions originate from individuals' situational assessments in relation to their objectives (as noted in Section \ref{sec:rq1.5}).

In developing sensor-free detection models, the selected studies utilized advanced data mining techniques, including feature selection approaches, resampling strategies, and robust strategies for hyperparameter tuning and evaluation such as grid search and k-fold cross-validation. These efforts enhance model performance and generability. The typical approach involves starting with a set of algorithms, then selecting a model based on a specific goal, usually superior performance according to one or more metrics. The most commonly used metrics are Cohen's Kappa and AUC.

The studies employed diverse algorithms to construct detectors, classified into Regression, Bayesian, Decision Tree, Ensemble-based, Instance-based, and Neural Networks. Despite the difficulties in direct comparisons due to the varying CBLEs, feature considerations, and emotion annotation methods, neural networks and decision tree algorithms generally yield the best results (as depicted in Figure \ref{fig:best_models}).

Regarding features, prominent features for detecting specific emotions were identified during feature engineering. Boredom was detected by considering features such as the number of consecutive incorrect answers and their speed. Confusion was linked to task types, task difficulty, the number of hints requested, and click patterns, among others. Delight was related to the number of gamification trophies, completed tasks, correct answers, and the time spent on various actions. Engagement and frustration were deduced from features including the number of completed tasks or correct answers, action history, the number of hints requested, and inactivity periods.

Most sensor-free emotion detection research involves K–12 students in the U.S., utilizing CBLEs in school labs for emotion data collection. Most studies have data from several hundred students, with only two studies collecting data from over 1,000 students. \cite{jensen2019generalizability} noted models trained with fewer than 1500 students lacked stable scores or predictions.

A significant proportion of the research was conducted using data from intelligent tutoring systems, reflecting the research community's strong background in artificial intelligence in education. The publication venues predominantly belong to the same field.

On model generalization, \cite{ocumpaugh2014population} noted that models are only applicable to different populations if trained with samples from the targeted population. As for generalizing models to other environments, existing results are preliminary.

Although current results exhibit the field's maturity, reflected in the robust methods of development and data collection, the area remains in the developmental stage. Only one study mentioned integrating the model with a CBLE, indicating the need for more studies on the performance of sensor-free detection models in real settings.

\section{Future Research Directions}

This systematic literature review has highlighted the significant progress made in the field of sensor-free affect detection in CBLEs. However, it also underscores several areas that warrant further exploration. These areas can be broadly categorized into three types: improving model performance, enhancing model development practices and methods, and integrating models into CBLEs.

\begin{enumerate}
    \item \textbf{Improving Model Performance}: The current models have shown promise in detecting students' emotions in CBLEs. They have utilized a range of machine learning techniques and have been trained on various types of student interaction data. However, there is room for improvement in their performance. Future research directions in this area could include:

    \begin{itemize}
        \item \textbf{Enhancing the Performance of Sensor-Free Detection Models}: Current models have demonstrated potential, yet there remains scope for performance enhancement. A key area for exploration is the potential improvement of model performance through the consideration of entire sequences of emotional labels and the application of recurrent neural networks. While advanced neural network algorithms have been explored in some studies, such as \citep{botelho2017improving,botelho2019machine}, these works utilized emotion labels collected via the BROMP protocol. This protocol annotates emotions in a round-robin fashion, moving to the next student following an annotation. Consequently, these studies do not capture the sequence of emotions experienced by individual students. Given that research indicates the existence of an emotional dynamic among students \citep{dmello2012dynamics}, incorporating this dynamic into the models could potentially enhance their emotion detection capabilities.

        \item \textbf{Collect More Samples of Underrepresented Emotions}: The research has primarily focused on widely studied learning-associated emotions such as engagement, confusion, frustration, and boredom. However, there is a need to collect more samples of underrepresented emotions like confusion and frustration. This will not only improve the accuracy of the models but also provide a more comprehensive understanding of students' emotional experiences in CBLEs.

        \item \textbf{Ascertain Other Emotions}: The current literature has primarily identified four affective states. Future research could explore whether other affective states might vary according to the CBLE. This could involve investigating whether there are other emotions beyond those mentioned in the literature and whether their frequency changes depending on the environment type.
   
\end{itemize}

\item \textbf{Enhancing Model Development Practices and Methods}: The development of these models has primarily been led by a small cluster of research groups, with a heavy reliance on data collection methods centered on online classroom observation. However, there is a need to refine these practices and methods. Future research directions in this area could include:

\begin{itemize}
    \item \textbf{Compare the Accuracy of Various Data Collection Techniques}: The current body of research is primarily led by a small cluster of research groups, with a heavy reliance on data collection methods centered on online classroom observation. Future research should compare the accuracy of various data collection techniques to identify the most effective methods for different learning environments and contexts.

    \item \textbf{Determine the Ideal Granularity of Duration}: Determining the ideal granularity of duration and assessing whether this parameter depends on the type of CBLE is crucial. Future research should explore if a better grain level exists and whether it depends on the learners or CBLEs.

    \item \textbf{Shared Database of Action Logs and Emotion Labels}: An emerging trend of note is open-science research, a movement aimed at reforming research practices to increase transparency and participation. This includes Open Access to publications, sharing of research data, transparency in research methods and processes, new means of research evaluation, and a reorientation of research to be more inclusive and responsive to societal and industrial needs \citep{RossHellauer2022}. The field of sensor-free affect detection could significantly benefit from a shared database of action logs and emotion labels. Such a resource would enable researchers to develop and compare their models using the same dataset. While initiatives like the LearnLab DataShop (\url{https://pslcdatashop.web.cmu.edu/}) exist within the realm of intelligent tutoring systems, to the best of our knowledge, no open database is currently available specifically for sensor-free emotion detection in CBLEs. Future research should prioritize exploring ways to overcome the associated challenges and make this valuable resource a reality.

    \item \textbf{Open Source Code}: In addition to a shared database, making the source code of these models publicly available in libraries would allow for greater transparency, reproducibility, and collaboration in the field. Future research should consider ways to facilitate this.
    \end{itemize}

\item \textbf{Integrating Models into CBLEs}: The practical application of these models in real learning environments is a crucial next step. This would allow for immediate intervention and adaptation based on the detected emotions, potentially optimizing learning outcomes. Future research directions in this area could include:

\begin{itemize}
    \item \textbf{Integrate Models into CBLEs for Real-Time Detection}: The practical application of these models in real learning environments is a crucial next step. Future research could focus on integrating these models into CBLEs and performing real-time detection of students' emotions. This would allow for immediate intervention and adaptation based on the detected emotions, potentially optimizing learning outcomes. In our SLR, we were able to find a unique work that integrated the model into a CBLE. 

    \item \textbf{Provide Meaningful Interventions Based on Detected Emotions}: Once emotions are detected in real-time, the next challenge is to provide meaningful interventions. Future research could explore how to best respond to detected emotions to enhance the learning experience. This could involve developing adaptive learning strategies or personalized feedback mechanisms based on the detected emotions.

    \item \textbf{Understand the Impact of Emotions on Learning}: While it is well-documented that emotions influence learning, there is still much to understand about this relationship. Future research could delve deeper into how specific emotions impact different aspects of learning, such as engagement, motivation, and learning outcomes. This could involve longitudinal studies or more nuanced analyses of student interaction data.
    \end{itemize}
\end{enumerate}

In conclusion, while substantial progress has been made in the field of sensor-free affect detection, there are numerous avenues for future research. By addressing these areas, we can move closer to the goal of creating CBLEs that are truly responsive to students' emotional states, thereby optimizing the learning process.

\section{Limitations and Threats to validity}

This systematic review, while comprehensive, has several limitations that should be acknowledged. 

First, the scope of the review was limited to studies published in English. This language restriction may have led to the omission of relevant studies published in other languages, potentially introducing a language bias and limiting the comprehensiveness of the review.

Second, the review relied on specific databases for the literature search. While these databases are widely used and respected, there may be relevant studies in other databases or grey literature that were not included in this review. This could potentially limit the breadth of the review and introduce a selection bias.

Third, the review process was subject to potential reviewer bias. Despite efforts to minimize this through independent reviews and consensus discussions, the subjective nature of study selection and data extraction processes may have influenced the results of the review.

Fourth, the review did not conduct a formal assessment of the quality or risk of bias in the individual studies included. While this is common in systematic reviews, it does limit the ability to assess the validity and reliability of the findings of the individual studies.

Fifth, the review did not include a meta-analysis due to the heterogeneity of the studies in terms of methodologies and measures used. This limits the ability to quantitatively synthesize the results and draw definitive conclusions.

Lastly, the field of sensor-free affect detection is rapidly evolving, with new studies continually being published. While this review offers a comprehensive overview of the field's status as of December 2021, the dynamic nature of this research area means that the content may soon be superseded by new findings and advancements.

In conclusion, while this systematic review provides valuable insights into the field of sensor-free affect detection, these limitations should be considered when interpreting the findings. Future systematic reviews could address these limitations by including studies in other languages, searching additional databases, conducting a formal quality assessment, and regularly updating the review to include new research.

\section*{Acknowledgment}

We would like to express our gratitude to the reviewers for their insightful comments and suggestions. Their expertise and dedication have significantly improved the quality of this manuscript. We also extend our sincere thanks to the editor for their guidance and support throughout the review process. 

Furthermore, we would like to acknowledge the support provided by our research funding bodies: ``blind review''.

\section*{Declaration of Generative AI Software tools in the writing process}

During the creation of this work, the authors utilized the AI language model ChatGPT 3.5 to proofread the entire manuscript, using the command "Proofread the latex text below, improving style, clarity and flow:". This operation was enabled by the Google Chrome Extension, editGPT. After employing this tool/service, the authors rigorously reviewed and revised the content as required. It is important to underscore that the authors accept complete responsibility for the final content of the publication.

\bibliographystyle{unsrtnat}
\bibliography{paper}

\end{document}